\documentclass[traditabstract]{aa} 
\usepackage{amssymb,amsmath}
\usepackage{subfig}
\usepackage{graphicx}
\usepackage{natbib}
\bibpunct{(}{)}{;}{a}{}{,} 
\usepackage{txfonts}
\usepackage{multirow}

\begin{document}

\title{The strongest gravitational lenses: I. The statistical impact of cluster mergers}

\titlerunning{Strong-lensing statistics and cluster mergers}

\author{M. Redlich \inst{1}, M. Bartelmann \inst{1}, J.-C. Waizmann \inst{2,3,4}, C. Fedeli \inst{5}}

\institute{Zentrum f\"ur Astronomie der Universit\"at Heidelberg, Institut f\"ur Theoretische Astrophysik, Albert-Ueberle-Str.~2, 69120 Heidelberg, Germany\\ \email{matthias.redlich@stud.uni-heidelberg.de}
\and
Dipartimento di Astronomia, Universit\`{a} di Bologna, via Ranzani 1, 40127 Bologna, Italy
\and
INAF - Osservatorio Astronomico di Bologna, via Ranzani 1, 40127 Bologna, Italy
\and
INFN, Sezione di Bologna, viale Berti Pichat 6/2, 40127 Bologna, Italy
\and
Department of Astronomy, University of Florida, 211 Bryant Space Science Center, Gainesville, FL 32611, USA
}

\authorrunning{M. Redlich et al.}

\date{\emph{A\&A manuscript, version \today}}

\abstract{For more than a decade now, it has been controversial whether or not the high rate of giant gravitational arcs and the largest observed Einstein radii are consistent with the standard cosmological model. Recent studies indicate that mergers provide an efficient mechanism to substantially increase the strong-lensing efficiency of individual clusters. Based on purely semi-analytic methods, we investigated the statistical impact of cluster mergers on the distribution of the largest Einstein radii and the optical depth for giant gravitational arcs of selected cluster samples. Analysing representative all-sky realizations of clusters at redshifts $z < 1$ and assuming a constant source redshift of $z_{\mathrm{s}} = 2.0$, we find that mergers increase the number of Einstein radii above $10 \arcsec$ $\left( 20 \arcsec \right)$ by $\sim 35 \%$ $\left( \sim 55 \% \right)$. Exploiting the tight correlation between Einstein radii and lensing cross sections, we infer that the optical depth for giant gravitational arcs with a length-to-width ratio $\ge 7.5$ of those clusters with Einstein radii above $10 \arcsec$ $\left( 20 \arcsec \right)$ increases by $\sim 45 \%$  $\left( \sim 85 \% \right)$. Our findings suggest that cluster mergers significantly influence in particular the statistical lensing properties of the strongest gravitational lenses. We conclude that semi-analytic studies must inevitably take these events into account before questioning the standard cosmological model on the basis of the largest observed Einstein radii and the statistics of giant gravitational arcs.}

\keywords{Cosmology: theory $-$ Gravitational lensing: strong $-$ Galaxies: clusters: general $-$ Methods: statistical}

\maketitle
%

\section{Introduction}

Both the abundance of gravitational arcs and the distribution of Einstein radii in galaxy clusters are valuable cosmological probes \citep{2010CQGra..27w3001B}. Therefore, the result of \citet{1998A&A...330....1B}, who reported that we observe ten times as many giant gravitational arcs on the sky as theoretically expected, poses a serious challenge. Various aspects of strong gravitational lensing that could potentially mitigate the tension between theory and observations were studied in a long series of subsequent works (see e.g. \citet{2008A&A...486...35F} or \citet[Sect. 5.2]{2010CQGra..27w3001B} for summaries, and references therein). In particular, \citet{2004MNRAS.349..476T} analysed numerically simulated mergers of galaxy clusters. These authors found that mergers substantially change the shape of the critical curves  and can boost a cluster's efficiency to produce giant arcs by an order of magnitude. \citet{2006A&A...447..419F} employed semi-analytic methods to estimate that cluster mergers approximately double the statistical strong-lensing efficiency of clusters at redshifts $z > 0.5$. Fedeli and coworkers argued that mergers might possibly explain the excess of gravitational arcs in observed galaxy clusters at moderate and high redshifts. However, \citet{2006A&A...447..419F} made several simplifying assumptions that we revise here: First, galaxy clusters were described by elliptically distorted spherical lens models instead of adopting more realistic triaxial density profiles \citep{2002ApJ...574..538J, 2003ApJ...599....7O}. This approximation reduces the required computing time substantially, since the calculation of deflection angles for triaxial density profiles involves numerical integrations \citep{1990A&A...231...19S}, while simple analytic expressions exist in the case of elliptically distorted density profiles \citep{1992grle.book.....S}. Second, all mergers were simulated with a fixed direction of motion and relative orientation of the merging clusters, neglecting two important degrees of freedom (see Section \ref{sec:relative_orientations}).

Recent studies indicate that the distribution of Einstein radii might also be in conflict with theory. More precisely, the largest observed Einstein radii \citep[e.g.][]{2008A&A...481...65H, 2008ApJ...684..177U, 2011MNRAS.410.1939Z} were claimed to exceed the maximum possible expectations of the standard cosmological model \citep{2008MNRAS.390.1647B, 2009MNRAS.392..930O, 2011A&A...530A..17M}. These conclusions were drawn by either comparing the largest observed Einstein radii to those found in numerical simulations or by semi-analytically estimating the probability of finding the strongest observed lens systems in a $\Lambda$CDM universe. While studies of Einstein radii in numerical simulations are probably most realistic, they always suffer from a limited sample size. The simulated boxes might simply be too small to contain a sufficient number of extraordinarily strong gravitational lenses, which forbids solid statistical conclusions. In a follow-up paper, we will show that this limitation is indeed decisive in the context of extreme value statistics \citep{2012arXiv1207.0801W}. Semi-analytic methods -- admittedly based on a set of simplifying assumptions -- can overcome this limitation because they are computationally less demanding and hence can be used to analyse large samples of particularly strong gravitational lenses within a comparably short time. However, we note that so far all semi-analytic studies of cosmological distributions of Einstein radii have only considered samples of isolated galaxy clusters. One important goal of this paper is to extend these previous approaches and to present a new semi-analytic method for studying distributions of Einstein radii that incorporates the impact of cluster mergers.

New findings of \citet{2011A&A...530A..17M} suggest that the excess of giant arcs and the too large Einstein radii are closely related. Analysing selected samples of strong gravitational lenses in the \textsc{MareNostrum} simulation \citep{2007ApJ...664..117G}, Meneghetti and collaborators discovered a remarkably tight correlation between lensing cross sections and Einstein radii of cluster-sized dark matter haloes. This correlation plays an important role for the structure and line of reasoning of this work.

This paper is structured as follows. Section \ref{sec:triaxial_lenses} summarizes an analytic model of triaxial gravitational lenses. In Section \ref{sec:lcs}, we compare three different methods for computing lensing cross sections for giant gravitational arcs and show how the alternative algorithms need to be adjusted so that they yield equally reliable results. Having identified a reliable measure of lensing cross sections, we first investigate the correlation between Einstein radii and lensing cross sections in Section \ref{sec:correlation_er_lcs} and then study how the correlation evolves during cluster mergers in Section \ref{sec:evolution_slp_merger}. Finally, in Section \ref{sec:statistics_er}, we introduce a semi-analytic method for computing cosmological distributions of Einstein radii that properly includes cluster mergers. Our conclusions are presented in Section \ref{sec:conclusions}.

In this paper, we consider two different sets of cosmological parameters obtained from the Wilkinson Microwave Anisotropy Probe (WMAP). These are the best-fitting parameters from the WMAP one-year \citep[WMAP1;][]{2003ApJS..148..175S}, $(\Omega_{\rm m0}, \Omega_{\Lambda 0}, \Omega_{\rm b0}, h, \sigma_8) = (0.7, 0.3, 0.045, 0.7, 0.9)$ and from the seven-year data \citep[WMAP7;][]{2011ApJS..192...16L}, $(\Omega_{\rm m0}, \Omega_{\Lambda 0}, \Omega_{\rm b0}, h, \sigma_8) = (0.727, 0.273, 0.0455, 0.704, 0.811)$.


\section{Triaxial gravitational lenses}
\label{sec:triaxial_lenses}

\subsection{The density profile of triaxial dark matter haloes}
\label{sec:tdmh}

\citet[][hereafter JS02]{2002ApJ...574..538J} performed a combined analysis of triaxiality in twelve high-resolution dark matter halo simulations to study the detailed shape of single haloes as well as haloes embedded in large cosmological simulations, to gather statistical information about halo profiles. These authors found that the universal, spherical density profile discovered by \citet[][hereafter NFW]{1996ApJ...462..563N} can be generalized to a triaxial model and showed that this generalization significantly improves the fit to simulated haloes. Moreover, by analysing large cluster populations in their cosmological simulations, JS02 derived probability density functions for the profile concentrations and axis ratios of triaxial dark matter haloes. The statistical description provided in JS02 allows constructing random catalogues of triaxial dark matter haloes that resemble realistic cosmological populations in numerical simulations.

\citet{2002ApJ...574..538J} parametrized the spatial density profile of a triaxial dark matter halo by means of Cartesian coordinates $\boldsymbol{x'} = (x',y',z')$ in the principal axis frame. Using this parametrisation, they proposed a generalization of the NFW density profile
\begin{align}
\label{eq:density_profile}
\rho(R) &= \frac{\delta_{\mathrm{ce}} \; \rho_{\mathrm{crit}}(z)}{(R/R_0)^{\alpha}(1+R/R_0)^{3-\alpha}} \; , \\ 
\label{eq:density_profile_2}
R^2 &\equiv \frac{{x'}^2}{(a/c)^2} + \frac{{y'}^2}{(b/c)^2} + {z'}^2 \qquad \left( a \leq b \leq c\right) \; .
\end{align}
Here, $z$ denotes the halo's redshift, $R_0$ is the scale radius (cf. Eq. \eqref{eq:scale_radius}), $\delta_{\mathrm{ce}}$ is the characteristic density (cf. Eq. \eqref{eq:characteristic_density}) and $\rho_{\mathrm{crit}}(z)$ denotes the critical density of the universe. The exact numerical value of the inner slope $\alpha$ of the density profile is still being discussed. While NFW originally proposed $\alpha = 1.0$, other authors argued that steeper profiles with values up to $\alpha = 1.5$ provide a better fit to observations and numerical simulations \citep{1999MNRAS.310.1147M, 2000ApJ...529L..69J, 2003MNRAS.338...14P, 2004MNRAS.349.1039N, 2008A&A...489...23L}. In contrast, recent combined strong- and weak-lensing analyses of selected clusters indicate shallower mass profiles with inner slopes $\alpha < 1$ \citep[see][for instance]{2011ApJ...728L..39N}. Following \citet{2003ApJ...599....7O}, we consider both $\alpha = 1$ and $\alpha = 1.5$ to cover a broad range of the predicted values and discuss some consequences of varying inner slopes on strong-lensing statistics. 

To draw a random triaxial dark matter halo of virial mass $M$ at redshift $z$, we first sample its axis ratios using the empirically derived probability density functions from JS02,
\begin{align}
\label{eq:p_ac}
p(a/c) &= \frac{1}{\sqrt{2\pi}\sigma_{\rm s}} \exp\left[-\frac{(a_{\mathrm{sc}}-0.54)^2}{2\sigma_{\mathrm{s}}^2}\right]\frac{\mathrm{d}a_{\mathrm{sc}}}{\mathrm{d}(a/c)} \; , \\ 
a_{\mathrm{sc}} &= \frac{a}{c} \left( \frac{M}{M_*} \right)^{0.07[\Omega_{\rm m}(z)]^{0.7}} \; , \\
p(a/b|a/c)  &=
\begin{cases} \dfrac{3}{2(1-r_{\mathrm{min}})}\left[1-\left(\dfrac{2a/b-1-r_{\mathrm{min}}}{1-r_{\mathrm{min}}}\right)^2\right] & a/b \ge r_{\mathrm{min}}\\ 0 & a/b < r_{\mathrm{min}} \end{cases} \; , \\ 
\label{eq:r_min}
r_{\mathrm{min}} &= \operatorname{max}\left(a/c,0.5\right) \; .
\end{align}
Here, $M_*$ is the characteristic nonlinear mass scale where $\sigma(M_*, z) = \delta_{\mathrm{c}}(z)$. According to JS02, the best-fitting parameter for the width of the axis ratio distribution $p(a/c)$ is $\sigma_{\mathrm{s}} = 0.113$.

The concentration $c_{\rm e}$ is defined by
\begin{equation}
c_{\rm e} \equiv \frac{R_{\rm e}}{R_0} \; ,
\end{equation}
where $R_{\mathrm{e}}$ is determined such that the mean density within the ellipsoid of the major axis radius $R_{\mathrm{e}}$ equals $\Delta_{\mathrm{e}}(z)\Omega(z)\rho_{\mathrm{crit}}(z)$, where
\begin{equation}
\Delta_{\rm e}(z) = 5\Delta_{\mathrm{vir}}(z) \left(\frac{c^2}{ab}\right)^{0.75} \; .
\end{equation}
$\Delta_{\mathrm{vir}}(z)$ is the overdensity of objects virialized at redshift $z$, which we approximate according to \citet{1997PThPh..97...49N}. JS02 found a log-normal distribution for the concentration, 
\begin{equation}
\label{eq:concentration_distribution}
p\left(c_{\mathrm{e}}\right) = \frac{1}{\sqrt{2\pi}\sigma_{\mathrm{c}}}\exp\left\{-\frac{\left[\operatorname{ln}(c_{\mathrm{e}})-\operatorname{ln}(\bar{c}_{\mathrm{e}})\right]^2}{2\sigma_{\rm c}}\right\}\frac{1}{c_{\mathrm{e}}} \; ,
\end{equation}
with a dispersion of $\sigma_{\rm c} = 0.3$. Following \citet{2003ApJ...599....7O}, we include a correlation between the axis ratio $a/c$ and the mean concentration,
\begin{align}
\label{eq:concentration}
\bar{c}_{\mathrm{e}} &= f_{\mathrm{c}} \, A_{\mathrm{e}} \sqrt{\frac{\Delta_{\mathrm{vir}}(z_{\rm c})}{\Delta_{\mathrm{vir}}(z)}} \left( \frac{1 + z_{\rm c}}{1 + z} \right) \; , \\
f_{\mathrm{c}} &= \mathrm{max} \left\{0.3, 1.35 \exp \left[ - \left(\frac{0.3}{a_{\mathrm{sc}}} \right)^2 \right] \right\} \; .
\label{eq:concentration_correction}
\end{align}
In Eq. \eqref{eq:concentration_correction}, we adopted a correction introduced by \citet{2009MNRAS.392..930O}, forcing $f_{\rm c} \ge 0.3$ to avoid unrealistically low concentrations for particularly low axis ratios $a_{\mathrm{sc}}$.  Additionally, as noted earlier by e.g. \citet{2004ApJ...610..663O}, triaxial haloes with particularly low axis ratios $a_{\mathrm{sc}}$ (and hence also low concentrations $c_{\rm e}$) are highly elongated objects whose lens potential is dominated by masses well outside the virial radius. While previous studies \citep[e.g.][]{2009JCAP...01..015B, 2009MNRAS.392..930O} tried to avoid these unrealistic scenarios by simply truncating the density profile (cf. Eq. \eqref{eq:density_profile}) outside the virial radius, we alternatively force all sampled axis ratios $a_{\mathrm{sc}}$ to lie within the range of two standard deviations. Since this manipulation may affect the statistics of the largest predicted Einstein radii, we will investigate the consequences in our follow-up work \citep{2012arXiv1207.0801W}. The numerical value of the free parameter $A_{\rm e}$ depends on the underlying cosmology. Throughout, we set $A_{\rm e} = 1.1$, which was proposed by JS02 for a standard $\Lambda$CDM model. The above expressions only apply to an inner slope of $\alpha = 1.0$. In the case of $\alpha = 1.5$, we used the simple relation $\bar{c}_{\rm e}(\alpha = 1.5) = 0.5 \times \bar{c}_{\rm e}(\alpha = 1.0)$ (\citet{2001ApJ...549L..25K}; JS02). $z_{\rm c}$ denotes the typical collapse redshift of a dark matter halo of mass $M$ and is computed using the complementary error function,
\begin{equation}
\label{eq:collapse_redshift}
\operatorname{erfc} \left\{ \frac{\omega(z_{\rm c})-\omega(0)}{\sqrt{2 \left[ \sigma^2(fM) - \sigma^2(M) \right]}} \right\} = \frac{1}{2} \; .
\end{equation}
Here, $\sigma^2(M)$ is the top-hat smoothed variance of the power-spectrum extrapolated to the redshift $z = 0$ and $\omega(z) = \delta_{\rm c}(z)/D(z)$, where $D(z)$ is the linear growth factor. Following JS02, we adopted $f = 0.01$. The typical collapse redshift can be derived within the framework of extended Press-Schechter theory \citep{1993MNRAS.262..627L}. It corresponds to the typical time when the most massive progenitor of a dark matter halo contained the mass fraction $fM$. However, note that JS02 defined the typical collapse redshift such that it does not depend on the halo's actual redshift.

All other profile parameters can be inferred from the axis ratios and the concentration. Using an empirical relation (found by JS02) between $R_{\rm e}$ and the spherical virial radius $r_{\mathrm{vir}}$, $R_{\rm e} / r_{\mathrm{vir}} \approx 0.45$,  
the scale radius $R_0$ is given by
\begin{equation}
\label{eq:scale_radius}
R_0 = 0.45 \, \frac{r_{\mathrm{vir}}}{c_{\rm e}} = \frac{0.45}{c_{\rm e}} \left[ \frac{3M}{4\pi\Delta_{\mathrm{vir}}(z)\Omega_{\rm m}(z)\rho_{\mathrm{crit}}(z)} \right]^{1/3} \; .
\end{equation}
Finally, the characteristic density $\delta_{\mathrm{ce}}$ of the density profile \eqref{eq:density_profile} is defined in terms of the concentration,
\begin{equation}
\label{eq:characteristic_density}
\delta_{\mathrm{ce}} = \frac{\Delta_{\rm e}(z)\Omega_{\rm m}(z)}{3}\frac{c_{\rm e}^3}{m(c_{\rm e})} \; ,
\end{equation}
where
\begin{equation}
m(c_{\rm e}) = 
\begin{cases}
\operatorname{ln}\left(1 + c_{\rm e}\right) - \dfrac{c_{\rm e}}{1 + c_{\rm e}}  & \quad (\alpha = 1.0) \\
 & \\
2\operatorname{ln}\left(\sqrt{c_{\rm e}} + \sqrt{1 + c_{\rm e}}\right) - 2 \sqrt{\dfrac{c_{\rm e}}{1 + c_{\rm e}}} & \quad (\alpha = 1.5)
\end{cases}
\; .
\end{equation}

\subsection{Basics of gravitational lensing}

Since the theory of gravitational lensing is well known, we abstain from giving a thorough introduction but instead refer to \citet{2010CQGra..27w3001B} for a recent review. We only summarize the quantities that are particularly important here.

Working in the thin-screen approximation, gravitational lensing essentially reduces to a mapping between the lens plane and the source plane. The lens mapping is given by the lens equation,
\begin{equation}
\boldsymbol{\beta} = \boldsymbol{\theta} - \boldsymbol{\alpha}\left(\boldsymbol{\theta}\right) \; ,
\end{equation}
which describes the relation between the position $\boldsymbol{\beta}$ of a source, the position $\boldsymbol{\theta}$ of the lensed image and the reduced deflection angle $\boldsymbol{\alpha}\left(\boldsymbol{\theta}\right)$. The magnification introduced by gravitational lensing is given by
\begin{equation}
\label{eq:lm_magnification}
\mu = \frac{1}{\lambda_{\rm t} \lambda_{\rm r}} \; ,
\end{equation}
where $\lambda_{\rm t}$ and $\lambda_{\rm r}$ denote the tangential and the radial eigenvalue of the lens mapping. Points on the lens plane where at least one eigenvalue vanishes are called \emph{critical points}. The set of critical points forms closed curves called \emph{critical curves}. Their images on the source plane are called \emph{caustics}. Equation \eqref{eq:lm_magnification} indicates that sources located close to caustics are strongly magnified by the lens mapping. Their images -- such as giant arcs -- appear in the regions close to the critical curves in the lens plane.

\subsection{Gravitational lensing by triaxial dark matter haloes}
\label{sec:lensing_tdmh}

Equation \eqref{eq:density_profile} defines the density profile of triaxial dark matter haloes in their principal axis frame. However, their lensing properties are determined by their projected surface mass density in the observer's frame. Since observed haloes may have arbitrary orientations and the required line-of-sight projection requires us to express the spatial density profile in terms of coordinates with respect to the observer, we first introduce an appropriate coordinate transformation.

Adopting the notation of Eq. \eqref{eq:density_profile}, $\boldsymbol{x'} = (x',y',z')$ denote the Cartesian coordinates in the principal axis frame of the halo, so that the $z'$-axis lies along the major axis of the ellipsoid. Moreover, we assume that the $z$-axis is aligned with the line-of-sight direction of the observer. Then, we can locally construct another Cartesian frame for the observer's coordinate system, denoted by $\boldsymbol{x} = (x,y,z)$. The origins of both coordinate systems are placed at the centre of the halo. \citet{2003ApJ...599....7O} exclusively investigated the lensing properties of individual triaxial haloes and hence could fix the relative orientation of the projected surface mass density profile in the $x$-$y$ plane at will. This freedom allowed them to use a simplified rotation matrix to parametrize the coordinate transformation between both frames. In Section \ref{sec:relative_orientations}, we show that the relative orientation of the projected mass density profiles of merging clusters plays an important role and accordingly we cannot adopt this practical simplification.  
\begin{figure}
	\centering
	\includegraphics[scale=0.3]{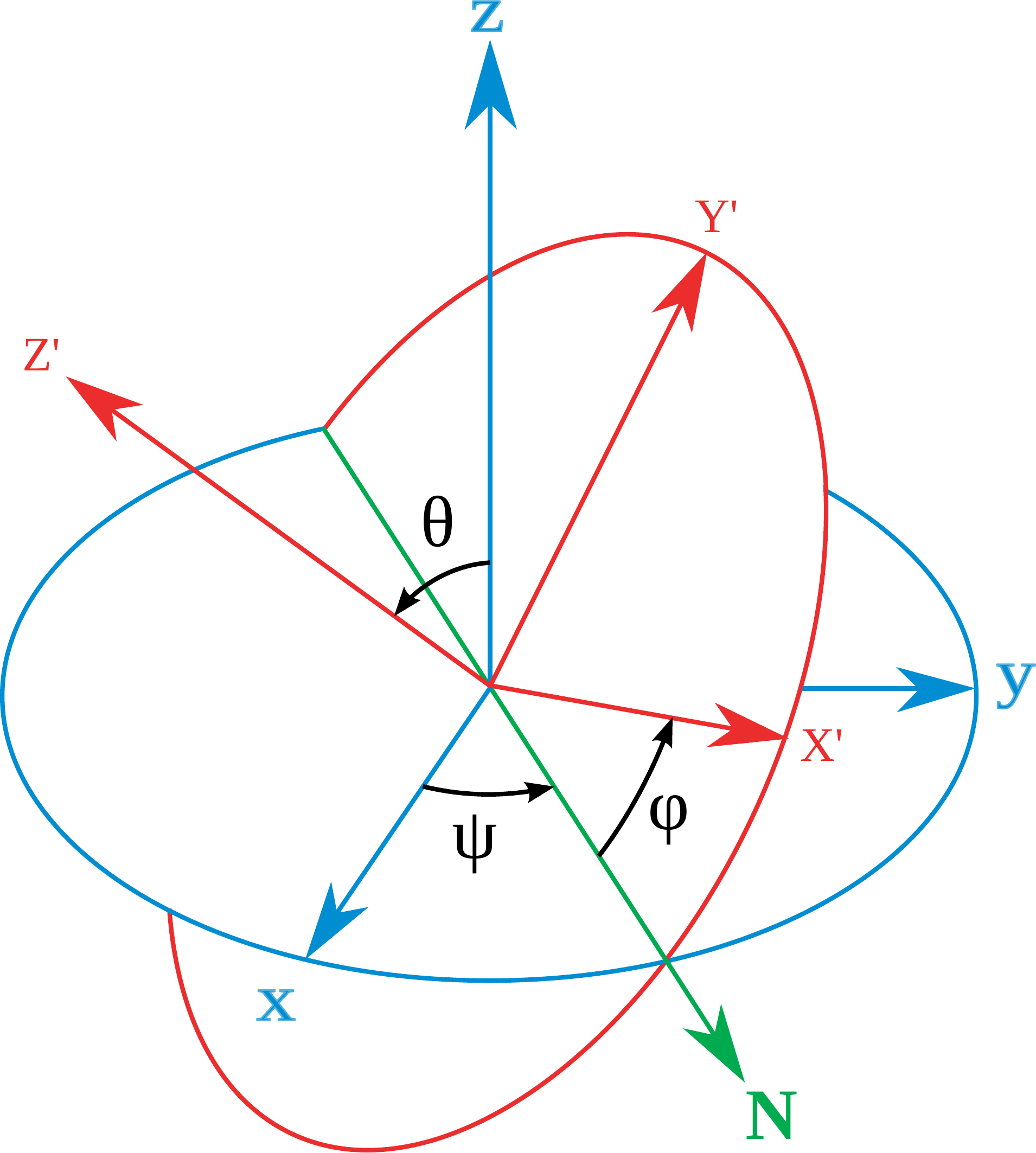}
	\caption[Visualization of the Euler angles used to parametrize a coordinate transformation.]{Visualisation of the Euler angles in the $z$-$x'$-$z''$ convention, which are used to parametrize the coordinate transformation between the observer's frame and the principal axis frame of the triaxial dark matter halo (figure adapted from Wikipedia (Lionel Brits)). }
	\label{fig:euler_angles}
\end{figure}
Instead we have to introduce a general coordinate transformation between both frames. To this end, we introduce the three Euler angles $(\psi, \theta, \phi)$ and adopt the so-called $z$-$x'$-$z''$-convention illustrated in Fig. \ref{fig:euler_angles}. The halo's coordinate system is rotated about the $z$-axis by the angle $\psi$, then about its new $x'$-axis by the angle $\theta$, and finally about the new $z''$-axis by the angle $\phi$. The coordinate transformation is given by
\begin{equation}
\boldsymbol{x'} = \boldsymbol{R}\boldsymbol{x} \; ,
\end{equation}
where the rotation matrix $\boldsymbol{R}$ is the product of the three rotation matrices about the axes.
Using this coordinate transformation, we can express the density profiles of arbitrarily oriented triaxial dark matter haloes in terms of the observer's coordinates. From here on, the derivation of all lensing properties of triaxial dark matter haloes follows \citet{2003ApJ...599....7O}, hence we refer the reader to their work. However, since we had to introduce a more general coordinate transformation, some algebraic coefficients are more complicated in our approach. We provide the necessary expressions in Appendix \ref{app:coordinate_transformation}. Moreover, we present a practical method to speed up the computation of deflection angles by a factor of $\sim 10$ in Appendix \ref{app:speed_up}.

In the following sections, we analyse strong-lensing properties of randomly drawn cosmological populations of triaxial dark matter haloes. Therefore we also need to sample random halo orientations. For that purpose, special care must be taken, since we cannot simply uniformly distribute the three angles $(\psi, \theta, \phi)$ in the corresponding angle bins. Instead, we have to apply the correct \emph{Haar measure} \citep{Haar1933}. Only $\psi$ and $\phi$ can be uniformly distributed in the range $[0, 2\pi]$. To sample $\theta$, we first draw a uniformly distributed random number $r$ in the range $[0,1)$ and then compute 
\begin{equation}
\theta = \operatorname{arccos}(1-2r) \; .
\end{equation}

\section{Computation of lensing cross sections}
\label{sec:lcs}

\subsection{Definition of the lensing cross section}

The strong-lensing efficiency of an arbitrary mass distribution is typically measured by the so-called lensing cross section $\sigma_p$, which is defined as the area of the region on the source plane where a source with given characteristics (morphology, orientation, surface brightness profile, etc.) has to lie in order to produce at least one image with properties $p$. This definition is quite general and needs to be additionally specified for the following applications. Throughout, we exclusively computed the lensing cross section $\sigma_{7.5}$, which measures the efficiency of a gravitational lens to produce highly elongated images (i.e. giant arcs) with length-to-width ratios exceeding $7.5$. For determining this area, each of the three following methods places random sources in the relevant regions of the source plane and analyses their lensed images. As was shown by e.g. \citet{1995A&A...297....1B} or \citet{2002ApJ...573...51O}, the size and the morphology of the background sources influence the resulting lensing cross section for gravitational arcs significantly, therefore a precise definition of these characteristics must be given. Following the common convention of previous works \citep[e.g.][] {1995A&A...297....1B, 2000MNRAS.314..338M, 2003MNRAS.340..105M, 2005A&A...442..405P}, we model the background sources as elliptical galaxies with constant surface brightness. Their size is chosen such that it equals the size of a circular source with a radius of $r = 0.5 \arcsec$. Their orientation (in the source plane) and minor-to-major axis ratios are randomly drawn from the ranges $[0,\pi]$ and $[0.5,1]$, respectively.

\subsection{Ray-tracing method A}

The first method for computing lensing cross sections was introduced by \citet{1993ApJ...403..497M}, adopted to nonanalytic lens models by \citet{1994A&A...287....1B} and further extended by subsequent works \citep{2000MNRAS.314..338M,2003MNRAS.340..105M,2005A&A...442..405P}. Elaborate descriptions of this method including all technical details are given there, hence we restrict ourselves to a brief summary of the aspects relevant for the following discussion.

\begin{figure}
	\centering
	\includegraphics[scale=0.5]{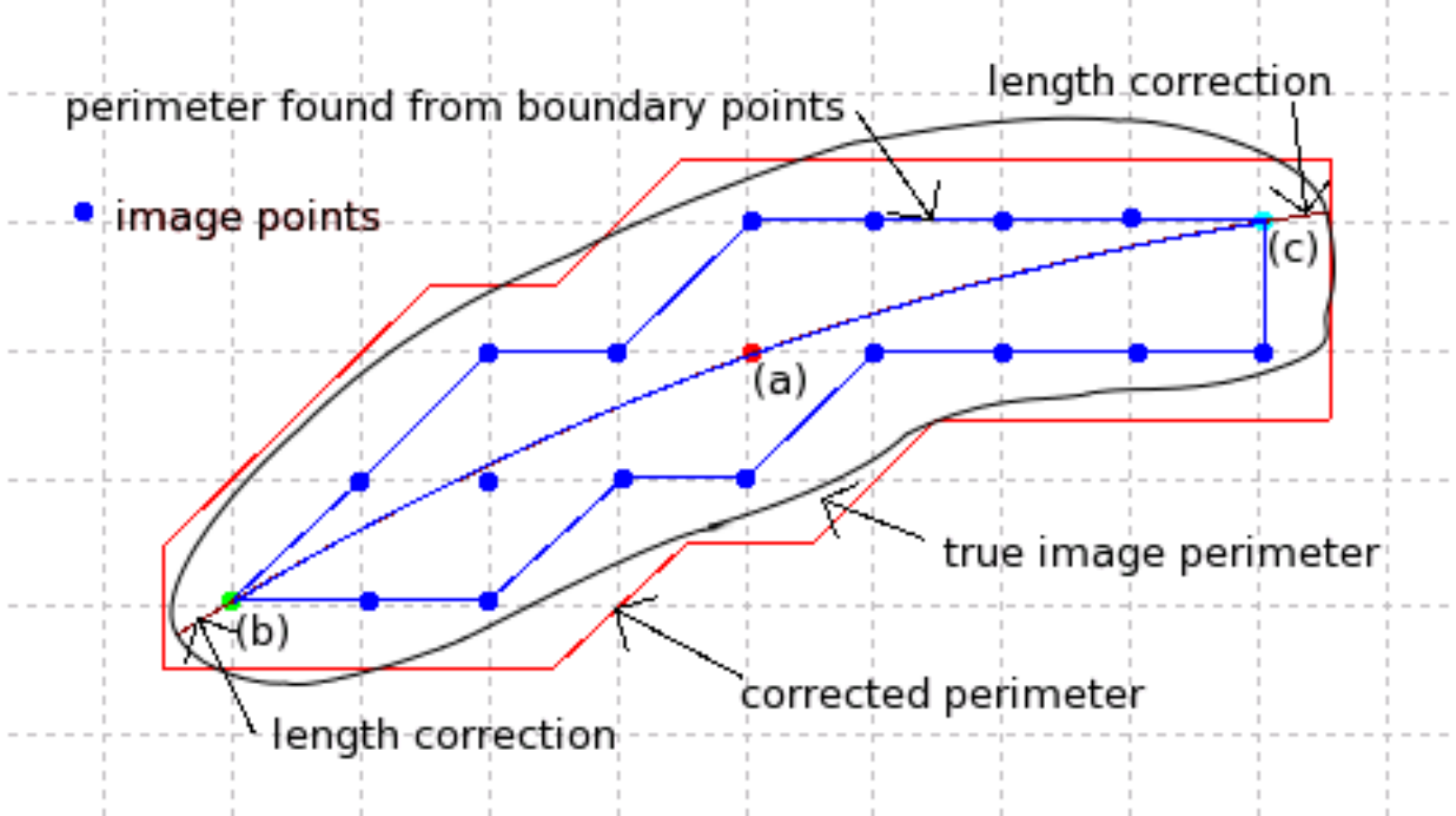}
	\caption[Illustration of the image-fitting algorithm of the ray-tracing method A.]{Illustration of the image fitting algorithm and the length- and perimeter-correction of the ray-tracing method A (figure from \citet{2005A&A...442..405P}).}
	\label{fig:ewald_image_fitting}
\end{figure}

First, the lens plane is covered with a Cartesian grid. At each grid point, the light deflection from gravitational lensing is computed to end up with a map of deflection angles. The resolution of this map must be sufficiently high so that the simulated images are resolved accurately and pixelisation effects are negligible (see below). Second, the source plane is covered with a Cartesian grid whose resolution is adaptively refined in regions close to the caustics. Next, the algorithm loops over each point of the refined grid and repeats the following set of operations. A random elliptical source is placed on the current grid point. Then, the images of this source are simulated using ray-tracing. A bundle of light rays starting at the observer is traced through the lens plane and mapped (by means of the lens equation) to the source plane. If a ray hits the source, the corresponding pixel on the lens plane is marked as an image point. After that procedure, the distinct resulting images are identified using a standard connected-component labelling algorithm and are separately analysed by a simple image-fitting algorithm, which is illustrated in Fig. \ref{fig:ewald_image_fitting}.
This algorithm sequentially determines three characteristic image points:

\begin{enumerate}
\item Point (a), which represents the arc centre and is given by the point whose image on the source plane falls nearest to the source centre.
\item Point (b) on the boundary, which is farthest away from the arc centre (a).
\item Point (c) on the boundary, which is farthest away from (b).
\end{enumerate}

After that, a circle is fitted through the points (a), (b), (c) and the length of the arc connecting these three points is preliminarily set as the image length $L$. Next, the perimeter is measured by walking along the ordered boundary points and summing up their mutual distances. Since the images are simulated using discrete grids, pixelisation effects have to be taken into account. \citet{2005A&A...442..405P} correctly noted that cells on the lens plane are only classified as belonging to the image if their centres fall within the source. Thus, boundary points are on average half a grid constant farther inside the image. To compensate for this effect, four grid constants must be added to the measured perimeter and the measured length must be elongated by one grid constant. The area $A$ of the image is found by simply counting the number of image pixels (blue points in Figure \ref{fig:ewald_image_fitting}) and summing up their corresponding cell sizes on the lens plane. Next, four different figures are fitted to the image:

\begin{enumerate}
\item A rectangle with area $A$, length $L$, and setting the width to $W = A/L$.
\item An ellipse with area $A$ and assuming that the length of the major axis equals the measured image length $L$. The image width $W$ is then defined as the length of the minor axis, $W = A/(\pi L)$.
\item A ring with outer radius $r_{\mathrm{out}} = L/2$ and choosing the inner radius $r_{\mathrm{in}}$ such that the area of the ring equals the measured area $A$ of the image. The image length is then redefined as $L = \pi (r_{\mathrm{out}} + r_{\mathrm{in}})$ and the width is set to the width of the ring, $W = r_{\mathrm{out}} - r_{\mathrm{in}}$.
\item A circle with radius $r = \sqrt{A/\pi}$ and hence defining $r = L = W$.
\end{enumerate}

After this fitting routine, the algorithm determines the best approximation of the real image shape by singling out the figure whose circumference agrees best with the measured image perimeter. If the length-to-width ratio of the chosen figure is higher than $7.5$, the corresponding cell size on the source plane is added to the lensing cross section $\sigma_{7.5}$. If the length-to-width ratios of multiple images of the considered source exceed that threshold, the cell size is multiplied by the corresponding image number. 

Performing the above procedure for all points of the adaptive grid on the source plane yields an accurate estimate of the lensing cross section $\sigma_{7.5}$.

\subsection{Ray-tracing method B}
\label{sec:rt_B}

\begin{figure}
	\centering
	\includegraphics[scale=0.4]{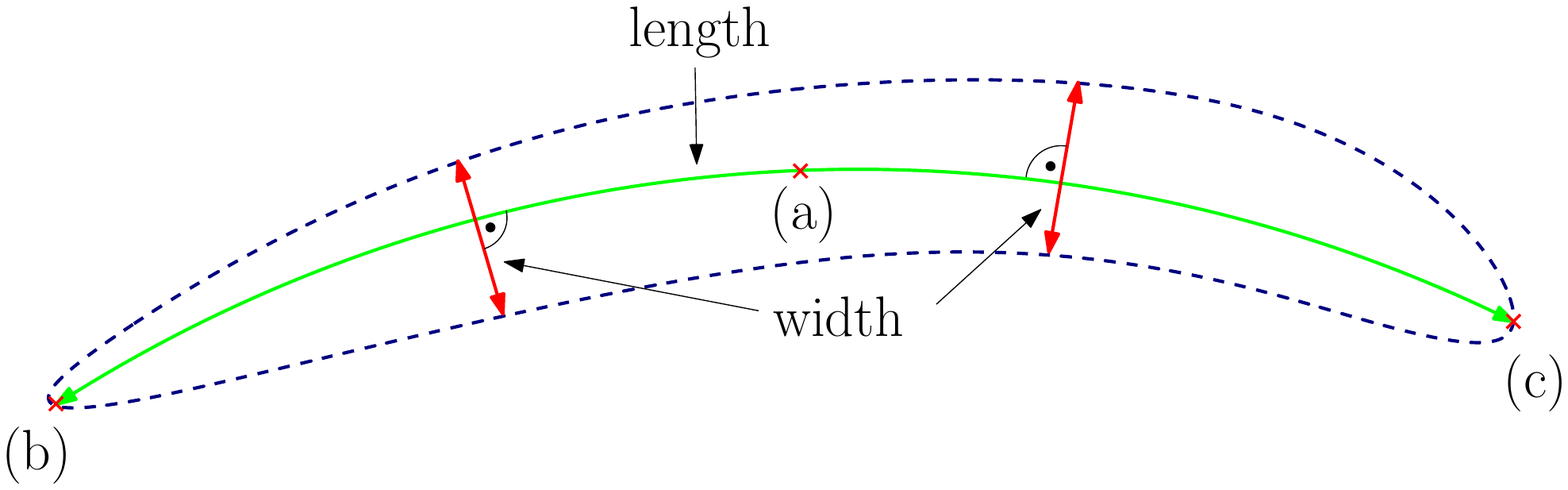}
	\caption[Illustration of the new algorithm of the ray-tracing method B to measure length-to-width ratios of gravitational arcs.]{Illustration of the new algorithm to measure length-to-width ratios of gravitational arcs. The image length is measured by fitting a circular arc (green arc) through the points (a), (b) and (c). Instead of fitting idealized, geometric figures to infer the width of the image, the average width is measured along the arc.}
	\label{fig:new_arc_fit}
\end{figure}

During a series of tests with method A, we observed that removing the length- and perimeter-correction introduced by \citet{2005A&A...442..405P} on average decreases the computed lensing cross sections by roughly $25 \%$, indicating that (unavoidable) errors in the length- and perimeter-measurements also have a significant impact on the results. Moreover, realistic shapes of gravitational arcs are rather irregular and only barely agree with the fitted, idealized geometric figures. To cross-check the results of method A, we developed a new ray-tracing method that employs an alternative algorithm to determine length-to-width ratios of lensed images. As in method A, randomly drawn elliptical sources are placed on an adaptively refined grid on the source plane. Again, the corresponding image configurations are found by tracing bundles of light rays through the lens plane. To identify the distinct lensed images, we employ the component-labelling algorithm proposed by \citet{Chang:2004:LCA:973388.973393}, since it returns a clockwise ordered list of boundary points by construction. This property is important for our alternative length-to-width measurement, which is illustrated in Fig. \ref{fig:new_arc_fit}. As with method A, we first identify the characteristic image points (a), (b), and (c) and fit a circle through them. Next, we determine the length of the (green) arc and correct for pixelisation effects by adding one grid constant. After that, we use points (b) and (c) to split the boundary into two parts. The first part consists of pixels connecting (b) and (c) in clockwise direction (upper part of the boundary) and the second group of pixels connects both points in anti-clockwise direction (lower part of the boundary). This separation of boundary points allows us to measure the average width of the image. For this, we first subdivide the angle that is enclosed by the arc connecting (b) and (c) into small bins. Then, we consider each angular bin separately, scan both boundaries for pixels that lie in the current range and use them to determine the image width at the considered position. Figure \ref{fig:new_arc_fit} visualizes this procedure for two specific angular bins along the arc. Finally, we define the \emph{average} of all measured widths as the width of the image. Again, we correct for pixelisation effects by adding one grid constant. In the same way as in method A, we determine the number of images whose length-to-width ratio is higher than $7.5$ for each source. We multiply this number with the corresponding cell size on the source plane and add this area to the final lensing cross section $\sigma_{7.5}$.

When we had completed this new algorithm, we realized that \citet{2008A&A...482..403M} used the same approach to measure the length-to-width ratio of arcs in realistic observations. But instead of \emph{averaging} the widths of the distinct angle bins, they computed the \emph{median} in order to be less sensitive to observational noise, which produces extreme outliers. Because we did not experience these problems in our purely theoretical computations, the simple average is expected to yield similar results.

\subsection{Semi-analytic method}

These ray-tracing simulations have a significant drawback: they can become computationally quite expensive. As already mentioned, the ray-tracing part requires computing a map of deflection angles. The resolution of this map must be sufficiently high so that pixelisation effects\footnote{The perimeter measurement is particularly sensitive to pixelisation effects. When images have an unfortunate orientation with respect to the Cartesian grid on the lens plane, their boundary, which should normally be smooth, is fairly rugged owing to their discrete representation on the grid. Thus, by summing the mutual distances between the ordered list of boundary points, the measured perimeter overestimates the correct circumference.} do not significantly falsify the measured length-to-width ratios. Performing a series of convergence tests, we found that the images of unmagnified circular sources with radius $r = 0.5\arcsec$ should at least be represented by $\sim 250$ pixels on the lens plane, which amounts to a grid constant of $\sim 0.056 \arcsec$. The deflection angle map needs to cover the entire region of the lens plane in which gravitational arcs can occur. Hence, for particularly strong gravitational lenses, the required side lengths can easily reach values of $\sim 200 \arcsec$ or more, which in turn lead to Cartesian grids with roughly $3500 \times 3500$ pixels or more. For triaxial gravitational lenses, the calculation of deflection angles involves numerical integrations and hence the computation of the deflection angle map can become rather expensive. Moreover, we note that the component-labelling algorithms, which are the most expensive operation of the ray-tracing methods, require more and more time as the resolution of the deflection angle map is increased because images are represented by ever larger ensembles of pixels. Finally, the resolution of the adaptive grid in the source plane must also be sufficiently high for a precise estimate of the lensing cross section. Consequently, many random sources must be placed and their images need to be simulated.

If arc statistics is meant to be used as a cosmological probe, lensing cross sections of large cluster samples need to be computed. In addition, these computations need to be repeated for different sets of cosmological parameters. Given the above requirements, this task quickly becomes computationally extremely demanding. To overcome this problem, \citet{2006A&A...447..419F} developed a fast and elegant method to calculate lensing cross sections semi-analytically. They found that their alternative method is faster than ray-tracing simulations by a factor of $\sim 30$ and yields equally reliable results. Since we aim to compare their method to the previously discussed ray-tracing simulations, a basic understanding of their new approach is required. Again, we only briefly outline the principal ideas and the most important formulae. We refer the reader to \citet{2006A&A...447..419F} for a thorough introduction including all technical details.

We note in the beginning that the intrinsic ellipticity of sources has a significant impact on the lensing cross section for giant gravitational arcs \citep[cf.][]{1995A&A...297....1B}. \citet{2001ApJ...562..160K} developed a simple and elegant formalism to treat this aspect mathematically, which we adopt as described by \citet{2006A&A...447..419F}. Without going into detail here, we assume that we possess an appropriate method for computing the average of the eigenvalue ratio $\langle q_{\rm l} \rangle$ $ \left( q_{\rm l} = max \left[ |\lambda_{\rm t}/\lambda_{\rm r}|, |\lambda_{\rm r}/\lambda_{\rm t}| \right] \right)$ over arbitrary images at arbitrary positions in the lens plane. Then, the area of the stripe in the lens plane where gravitational arcs with a length-to-width ratio equal to or higher than $7.5$ occur is simply given by
\begin{equation}
B_l =  \int H \left[ \langle {q}_{\mathrm{obs}} \rangle (\boldsymbol{\theta}) - 7.5 \right] \; \mathrm{d}^2\theta \; ,
\end{equation}
where $H(x)$ is the Heaviside step function and the integration is carried out over the lens plane. The notation $\langle q_{\mathrm{obs}} \rangle$ indicates that we use the average eigenvalue ratio $\langle q_{\rm l} \rangle$ for computing the observed axis ratio. Using the lens equation, $B_{\rm l}$ can be mapped to an area $B_{\rm s}$ on the source plane, which is by definition the lensing cross section $\sigma_{7.5}$ we are searching for. Taking the magnification caused by the lens mapping into account, the lensing cross section is thus given by
\begin{equation}
\label{eq:sa_lcs_int}
\sigma_{7.5} = \int H \left[ \langle q_{\mathrm{obs}} \rangle(\boldsymbol{\theta}) - 7.5 \right] |\operatorname{det}A\left(\boldsymbol{\theta}\right)| \; \mathrm{d}^2\theta \; ,
\end{equation}
where $\operatorname{det}A\left(\boldsymbol{\theta}\right)$ denotes the Jacobian determinant of the lens mapping. Effectively, this elegant approximation reduces the lensing cross section computation to a simple area integral, which can be evaluated numerically. 

\subsection{Comparison of the different methods}
\label{sec:lcs_comparison}

We are now in the position to carry out the comparison between the three different methods to check their reliability. To this end, we sampled 300 random triaxial haloes placed at redshift $z_{\rm l} = 0.5$ and computed their lensing cross sections $\sigma_{7.5}$ using each of the three algorithms. The halo masses were logarithmically uniformly distributed within $(5 \times 10^{14}  - 2 \times 10^{15}) \,  M_{\odot} h^{-1}$, and we adopted $\alpha = 1.0$ for the inner slope of their density profile. The source plane was placed at redshift $z_{\rm s} = 2.0$.
\begin{figure}
	\centering
	\includegraphics[scale=0.8]{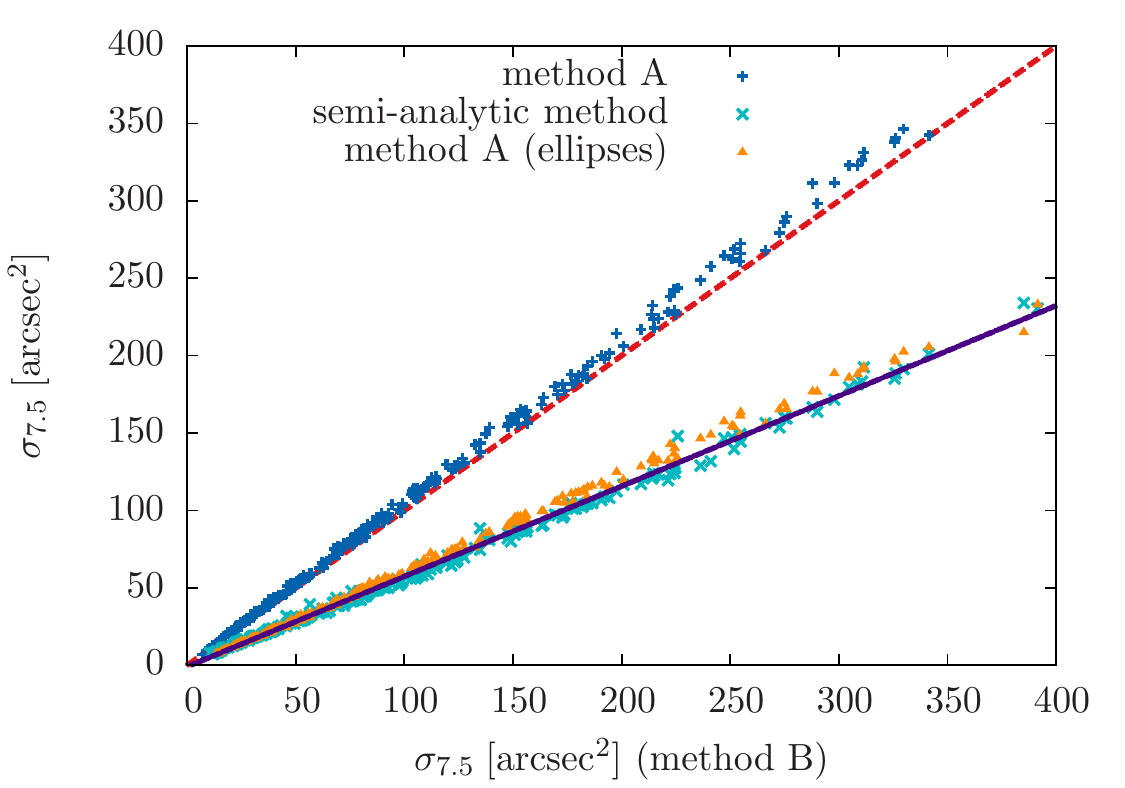}
	\caption[Lensing cross sections according to the ray-tracing simulations and the semi-analytic method.]{Lensing cross sections $\sigma_{7.5}$ in the range of $0-400 \, \mathrm{arcsec}^2$ according to method A (blue crosses), the semi-analytic method (turquoise crosses) and method A fitting ellipses only (yellow triangles) as a function of the lensing cross sections $\sigma_{7.5}$ computed with method B for a sample of 300 massive dark matter haloes (see text). The red dashed line indicates the bisector of the $\sigma_{7.5}$ - $\sigma_{7.5}$ plane. The violet dashed line represents the best linear fit of the semi-analytic lensing cross sections.}
	\label{fig:lcs_comparison}
\end{figure}

Figure \ref{fig:lcs_comparison} shows the lensing cross sections $\sigma_{7.5}$ computed with ray-tracing method A and the semi-analytic method as a function of the lensing cross sections $\sigma_{7.5}$ computed with ray-tracing method B. For reasons to be clarified below, we also plot the results according to method A but fitting only ellipses to the simulated images.

Obviously, both ray-tracing methods agree excellently in the range $\sigma_{7.5} \lesssim 225 \, \mathrm{arcsec}^2$, although they employ inherently different algorithms to measure length-to-width ratios of lensed images. For lensing cross sections $\sigma_{7.5} \gtrsim 225 \, \mathrm{arcsec}^2$, the results computed with method A tend to be slightly higher. However, we solely attribute these deviations to technical limitations. The implementation of method A was written in \textsc{Fortran-77} and we experienced problems when allocating large amounts of memory. Accordingly, we had to limit the maximum possible grid dimension for the deflection angle maps to $2048 \times 2048$ pixels. For the largest shown lensing cross sections ($\sigma_{7.5} \gtrsim 225 \, \mathrm{arcsec}^2$), the considered field sizes might simply be too large for the highest possible resolution, leading to coarsely resolved images. Hence the errors due to pixelisation effects increase and account for the differences between both ray-tracing simulations. Given these explanations, we conclude that both ray-tracing methods yield equally reliable results.

In contrast to \citet{2006A&A...447..419F}, however, we cannot reproduce the good agreement between the semi-analytic method and the ray-tracing simulations. Instead, the semi-analytic lensing cross sections are systematically too small. In order to quantify the deviations, we computed the best linear fit:
\begin{equation}
\label{eq:lcs_linear_fit}
\sigma_{7.5,\mathrm{SA}} = \left(0.583 \pm 0.002\right) \sigma_{7.5,\mathrm{B}} - \left(0.931 \pm 0.287\right) \; .
\end{equation}
Here, $\sigma_{7.5,\mathrm{SA}}$ and $\sigma_{7.5,\mathrm{B}}$ denote the lensing cross sections computed with the semi-analytic method and method B, respectively. In Fig. \ref{fig:lcs_comparison}, the above linear relation \eqref{eq:lcs_linear_fit} is indicated by the violet dashed line. Evidently the scatter around the linear relation is small, indicating a strong correlation between both quantities. The small fluctuations with respect to the linear fit can be explained by different realisations of the randomly drawn source populations: as shown by  \citet{2001ApJ...562..160K}, the observed length-to-width ratios of lensed images depend on the sources' intrinsic ellipticities and relative orientations with respect to the eigenvectors of the magnification tensor. Since these properties are randomly drawn when placing the sources on the adaptive grid, some cells only contribute to the lensing cross section if the source placed on top of them has favourable properties. Thus the lensing cross sections experience statistical fluctuations.

The above observations suggest that there must be a fundamental difference between the semi-analytic method and the ray-tracing simulations. As already mentioned, we additionally computed the lensing cross sections with a slightly modified version of method A. Instead of probing four different geometrical shapes, we fitted the simulated images with ellipses only. Obviously, this modified ray-tracing algorithm yields results that agree well with the semi-analytic method. This information turns out to be useful in the next section.

\subsection{Reconciling the semi-analytic method with the ray-tracing simulations}
\label{sec:lcs_reconcile}

The findings of the previous sections suggest a simple way for reconciling the semi-analytic method with the ray-tracing simulations. We recall that we found excellent agreement between both ray-tracing methods although their algorithms for determining the width of lensed images differ substantially. However, further analysis reveals that method A fits $\sim 99\%$ of the detected arcs with rectangles, which effectively equals method B's measuring the average image widths without imposing additional geometrical constraints.  Method A yields almost identical results to the semi-analytic method if only ellipses are fitted to the simulated images. These findings simply confirm that the semi-analytic method works as it should: By assuming that the averaged observable axis ratios $\langle q_{\mathrm{obs}} \rangle$ equal the length-to-width ratios of images, the method implicitly assumes that the lensed images roughly look like ellipses. Hence the good agreement with the modified version of method A is rather reassuring than indicating a failure of the semi-analytic approximations.

\begin{figure}
	\centering
	\includegraphics[scale=0.75]{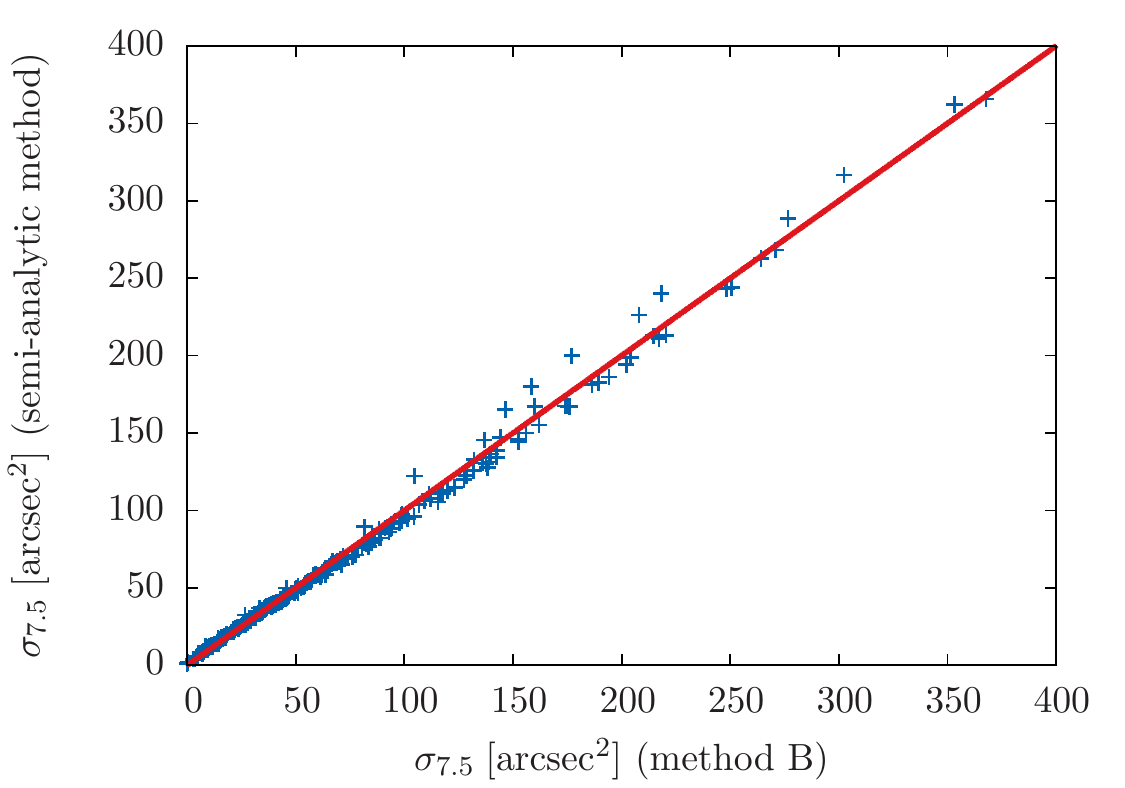}
	\caption[Figure comparing the modified semi-analytic method to the ray-tracing method B.]{Lensing cross sections $\sigma_{7.5}$ in the range of $0-400 \, \mathrm{arcsec}^2$ computed with the modified semi-analytic method as a function of the lensing cross sections $\sigma_{7.5}$ calculated with ray-tracing method B. The red dashed line indicates the bisector of the $\sigma_{7.5}$-$\sigma_{7.5}$ plane. The lensing cross sections were computed for a sample of 300 massive triaxial dark matter haloes (cf. Section \ref{sec:lcs_comparison}).}
	\label{fig:lcs_semi_analytic_modified}
\end{figure}

We thus find a simple solution reconciling the semi-analytic method with the ray-tracing simulations. We therefore consider an arbitrary image with area $A$ and length $L$. Fitting a rectangle, we first obtain the width $W = A/L$, and thus the length-to-width ratio is given by
\begin{equation}
\label{eq:ltw_rectangle}
\frac{L}{W} = \frac{L^2}{A} \; .
\end{equation}
Alternatively, we can also fit an ellipse. Using the following relation between the area of an ellipse, its length and its width,
\begin{equation}
A = \pi \frac{L}{2} \frac{W}{2} \; ,
\end{equation}
we obtain the length-to-width ratio
\begin{equation}
\label{eq:ltw_ellipse}
\frac{L}{W} = \frac{\pi}{4} \frac{L^2}{A} \; .
\end{equation}
Obviously, if we fix the area $A$ and the length $L$ (as both ray-tracing methods do), rectangles have higher length-to-width ratios than ellipses and the difference is simply given by a factor of $\pi/4$. That is precisely the reason why lensing cross sections decrease by $\sim 40\%$ if only ellipses are fitted to the simulated images. The measured length-to-width ratios are lower and therefore fewer images are recognized as giant arcs. On the other hand, this suggests an easy solution for adapting the semi-analytic method to the ray-tracing simulations. Recalling that the majority of arcs seems to be better approximated by rectangles, we can simply multiply the calculated observed axis ratio $q_{\mathrm{obs}}$ by $4/\pi$ to estimate the correct length-to-width ratio of most images. This simple multiplication is the mathematical equivalent of fitting images with rectangles instead of ellipses. To verify this idea, we sampled another random population of triaxial haloes following the same recipe as in the previous section and computed their lensing cross sections with the modified semi-analytic method and method B. Figure \ref{fig:lcs_semi_analytic_modified} clearly indicates that the agreement between both methods is now reasonably good and the deviations typically remain well below $15 \%$.

We therefore note that the significant and systematic deviations between the semi-analytic method and the ray-tracing simulations are simply caused by different definitions of length-to-width ratios of images. Gravitational arcs have irregular shapes and barely agree with idealized, geometrical figures, leaving no natural definition of a length-to-width ratio. Since Fig. \ref{fig:lcs_comparison} and Eqs. \eqref{eq:ltw_rectangle}-\eqref{eq:ltw_ellipse} clearly show that the actual definition has a significant impact on the final lensing cross section, it is particularly important to state precisely which convention is used to yield comparable results. Moreover, the semi-analytic method can be reconciled with the ray-tracing simulations by introducing a simple correction, so that it still provides a fast, reliable alternative for computationally demanding problems. Nevertheless, for the remainder of this paper, we always use ray-tracing method B to compute of lensing cross sections.

\section{The correlation between Einstein radii and lensing cross sections}
\label{sec:correlation_er_lcs}

\subsection{Two definitions of Einstein radii}

Einstein radii measure the size of the tangential critical curve. While a natural choice exists for the Einstein radius in the case of axially symmetric lenses, it is not obvious how this concept can be transferred to arbitrary lenses with irregular tangential critical curves. For this reason various alternatives were proposed by several authors \citep{2008MNRAS.390.1647B, 2009MNRAS.392..930O, 2011MNRAS.410.1939Z, 2011A&A...530A..17M}. The following two definitions turned out to be most useful for our purposes.

The first definition is of statistical nature. Let $\boldsymbol{\theta}_t$ denote the set of tangential critical points. The \emph{median Einstein radius} $\theta_{\mathrm{E,med}}$ is defined as the median distance of these points with respect to the lens' centre $\boldsymbol{\theta}_{\rm c}$,
\begin{equation}
\theta_{\mathrm{E,med}} \equiv \mathrm{median} \; \left\{ \sqrt{\left( \theta_{i,x} - \theta_{c,x} \right)^2 + \left( \theta_{i,y} - \theta_{c,y} \right)^2} \; \mid \; \boldsymbol{\theta}_i \in \boldsymbol{\theta}_t \right\} \; .
\end{equation}
This definition was introduced by \citet{2011A&A...530A..17M}, who analysed strong-lensing properties of numerically simulated clusters with irregular shapes that also showed substructure. The centres of those clusters were defined as the location of the maximum of the smoothed projected density distribution. We note that the \emph{median} was chosen in order to be less sensitive to extreme outliers of irregular critical curves.

The second definition is of geometrical nature. Let $A$ denote the area enclosed by the tangential critical curve. Then, the \emph{effective Einstein radius} $\theta_{\mathrm{E,eff}}$ is defined by
\begin{equation}
\theta_{\mathrm{E,eff}} \equiv \sqrt{\frac{A}{\pi}} \; ,
\end{equation}
such that a circle with radius $\theta_{\mathrm{E,eff}}$ has the area $A$. Evidently, the median Einstein radius and the effective Einstein radius both agree with the original definition for axially symmetric lenses.

\subsection{Correlation for numerical clusters}

\begin{figure}
	\centering
	\includegraphics*[viewport=20 10 500 350, scale=0.49]{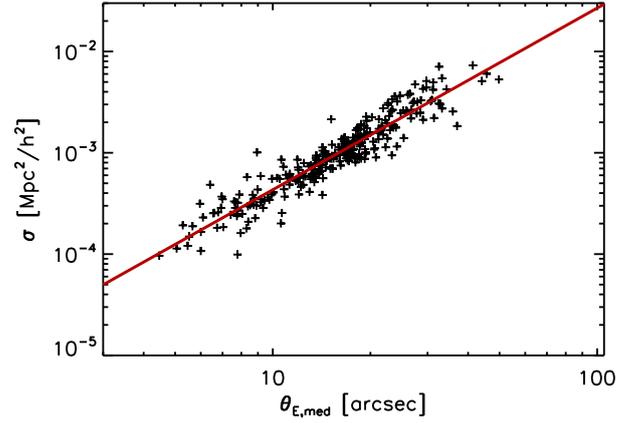}
	\caption[Correlation between Einstein radii and lensing cross sections in the \textsc{MareNostrum} simulation.]{Strong correlation between the median Einstein radii $\theta_{\mathrm{E,med}}$ and the strong-lensing cross sections $\sigma_{7.5}$ of a sample of massive clusters with masses $M > 5 \times 10^{14} M_{\odot} h^{-1}$ and redshifts $z_{\rm l} > 0.5$ extracted from the \textsc{MareNostrum Universe}. The red solid line indicates the best linear fit relation between $\log(\sigma_{7.5})$ and $\log(\theta_{\mathrm{E,med}})$ (figure from \citet{2011A&A...530A..17M}).}
	\label{fig:correlation_meneghetti}
\end{figure}
\citet{2011A&A...530A..17M} extracted a sample of massive clusters $\left( M > 5 \times 10^{14} M_{\odot} h^{-1} \right)$ at redshifts $z_{\rm l} > 0.5$ from the \textsc{MareNostrum Universe} \citep{2007ApJ...664..117G} and performed a statistical analysis of their strong-lensing properties. Figure \ref{fig:correlation_meneghetti} shows parts of their results, indicating a tight correlation between the lensing cross sections $\sigma_{7.5}$ and the median Einstein radii $\theta_{\mathrm{E,med}}$ of their cluster sample. Performing a linear fit to all data points in the $\log(\theta_{\mathrm{E,med}}) - \log(\sigma_{7.5})$ plane, they obtained
\begin{equation}
\label{eq:correlation_meneghetti}
\log (\sigma_{7.5}) = (1.79 \pm 0.04) \log(\theta_{\mathrm{E,med}}) - (5.16 \pm 0.05) \; ,
\end{equation}
with a Pearson correlation coefficient of $r = 0.94$, confirming the tight correlation between both quantities.

These findings are important for several reasons. First, as already noted by \citet{2011A&A...530A..17M}, they clearly indicate the strong connection between the problem of too large Einstein radii and the arc statistics problem. A particularly strong gravitational lens, whose Einstein radius is too large for the $\Lambda$CDM model, will most likely also have a lensing cross section that exceeds the maximum theoretical expectations. Hence, if we were to observe too large Einstein radii, we would also expect an excess of gravitational arcs. On the other hand, we note that the converse is not true and care must be taken, since both problems are not identical. Even if there were a way to explain the particularly large Einstein radii within the $\Lambda$CDM model, this would not necessarily solve the arc statistics problem. While the distribution of largest Einstein radii only tests the extreme cases, an arc statistic additionally measures the cumulative lensing efficiency of the entire halo population, so that it is also sensitive to the pure abundance of relatively unspectacular lenses, for instance. Thus, if the observed universe simply contained more strong gravitational lenses than theoretically predicted, we could hypothetically still observe an excess of gravitational arcs without having the problem of too large Einstein radii. 

Secondly, the correlation discovered might enable us to compute arc statistics for certain halo samples following a new approach. Even by means of the fast semi-analytic method, the calculation of lensing cross sections for large halo samples is a computationally demanding problem, since complete maps of deflection angles have to be computed for each individual lens. In comparison, the calculation of Einstein radii can be fast because only few deflection angles near critical curves need to be computed (see Section \ref{sec:computation_einstein_radii}). Hence, the idea is to sample cosmological distributions of dark matter haloes, compute their Einstein radii, and finally infer their lensing cross sections by means of the correlation. To this end, we need to investigate if we can reproduce the findings of \citet{2011A&A...530A..17M} with analytic mass profiles for triaxial haloes.

\subsection{Correlation for clusters with analytic mass profile}

We mimicked the cluster sample from \citet{2011A&A...530A..17M} by drawing random cosmological populations of massive $\left(M > 5 \times 10^{14} M_{\odot} h^{-1}\right)$ triaxial dark matter haloes at redshifts $z_{\rm l} > 0.5$ (cf. Appendix \ref{app:drawing_cosmological_populations}). To investigate the impact of variations of the cosmological parameters on the precise functional form of the correlation, we analysed cluster samples within both a WMAP1 and a WMAP7 cosmology. Furthermore, since the steepness of the inner core of dark matter density profiles is known to substantially change the strong-lensing properties of galaxy clusters \citep[see][]{2010CQGra..27w3001B}, we additionally considered the two extreme values of the inner slope, namely $\alpha = 1.0$ and $\alpha = 1.5$. 

\begin{figure}
	\centering
	\includegraphics[scale=0.95]{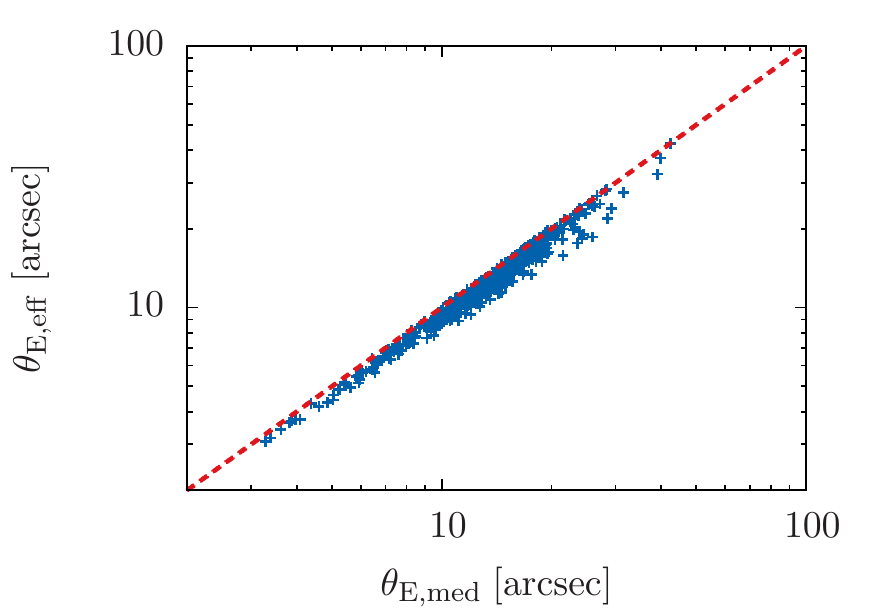}
	\caption[Comparison between effective and median Einstein radii for a cosmological population of triaxial dark matter haloes.]{Comparison between effective and median Einstein radii for a cosmological population of triaxial dark matter haloes with masses $M > 5 \times 10^{14} M_{\odot} h^{-1}$ at redshifts $z_{\rm l} > 0.5$, assuming a WMAP7 cosmology and adopting density profiles with inner slope $\alpha = 1.0$. Each cross represents a pair of Einstein radii. The red dashed line indicates the bisector of the $\log\left(\theta_{\mathrm{E,med}}\right)$-$\log\left(\theta_{\mathrm{E,eff}}\right)$ plane. The effective Einstein radii are systematically smaller, as is to be expected for ellipsoidal surface mass densities.}
	\label{fig:theta_E_eff_vs_theta_E_med}
\end{figure}

Before we analyse the discussed correlation for the different settings, we begin with a short comparison of the alternative definitions of Einstein radii. Figure \ref{fig:theta_E_eff_vs_theta_E_med} shows effective versus median Einstein radii for one specific population of dark matter haloes, namely assuming a WMAP7 cosmology and adopting triaxial density profiles with inner slope $\alpha = 1.0$. We note that these plots qualitatively look the same for all other realisations (WMAP1 cosmology or steeper density profiles). Obviously, the agreement between the two definitions is moderate, and the absolute deviations increase with larger Einstein radii. Moreover, almost all points lie below the bisector, indicating that effective Einstein radii are systematically smaller than median Einstein radii. However, this was to be expected, since both definitions only agree for axially symmetric lenses and start to differ $\left( \theta_{\mathrm{E,eff}} < \theta_{\mathrm{E,med}} \right)$ as soon as ellipticity is introduced. The deviation is especially pronounced when highly ellipsoidal surface mass density profiles are considered.

\begin{figure}
	\centering
	\includegraphics[scale=0.95]{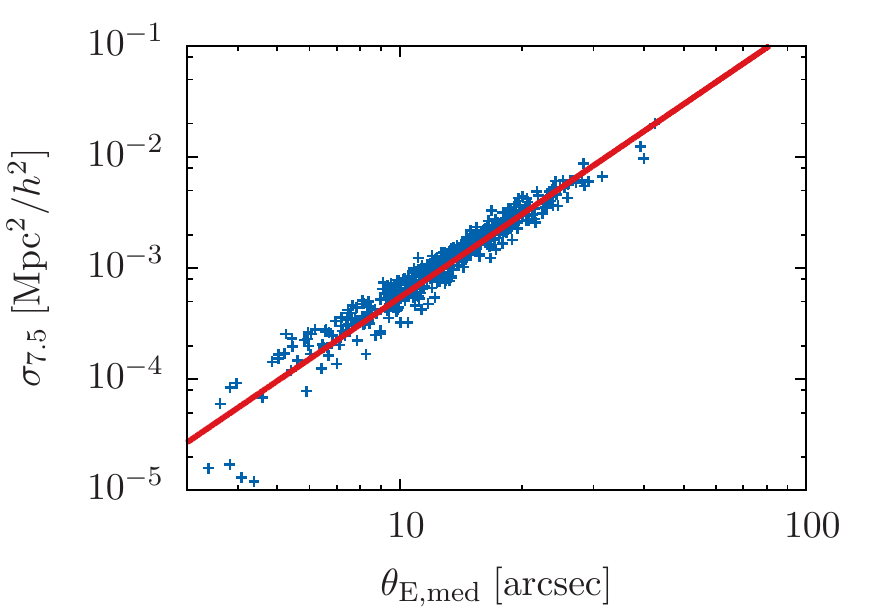}
	\caption[Correlation between median Einstein radii and lensing cross sections for a cosmological population of triaxial dark matter haloes.]{Strong correlation between median Einstein radii $\theta_{\mathrm{E,med}}$ and lensing cross sections $\sigma_{7.5}$ for a cosmological population of triaxial dark matter haloes $\left(M > 5 \times 10^{14} M_{\odot} h^{-1}\right)$ at redshifts $z_{\rm l} > 0.5$, adopting a density profile with inner slope $\alpha = 1.0$. The population was drawn assuming a WMAP7 cosmology. The red solid line indicates the best linear fit relation between $\log(\sigma_{7.5})$ and $\log(\theta_{\mathrm{E,med}})$. The correlation breaks down for small Einstein radii $\left( \theta_{\mathrm{E,med}} \lesssim 8 \arcsec \right)$.}
	\label{fig:theta_E_med_vs_lcs_alpha1}
\end{figure}

In Fig. \ref{fig:theta_E_med_vs_lcs_alpha1}, we plot lensing cross sections $\sigma_{7.5}$ as a function of median Einstein radii $\theta_{\mathrm{E,med}}$ for the same cluster sample as before. We can evidently reproduce the findings of \citet{2011A&A...530A..17M}. The correlation is especially tight for large Einstein radii, and thus also for large lensing cross sections. The scatter progressively increases as the size of the Einstein radii decreases until the correlation finally breaks down for Einstein radii $\theta_{\mathrm{E,med}} \lesssim 8\arcsec$. We recall that we considered elliptical sources with an equivalent radius of $r_s = 0.5 \arcsec$ to compute lensing cross sections. In the regime where the correlation breaks down, the size of these sources is roughly of the same order as the size of the caustics, so that the efficiency to produce highly elongated images rapidly drops. Hence, as soon as the considered Einstein radii become too small, the correlation cannot be used anymore to infer lensing cross sections. Of course this breakdown also occurs in the case of effective Einstein radii.

\begin{table*}
\caption[Best linear relation between Einstein radii and lensing cross sections for different cosmological populations of triaxial dark matter haloes.]{Results of the least-squares fits for the relation between Einstein radii and lensing cross sections, assuming a correlation of the form $\log\left(\sigma_{7.5}\right) = a\log\left(\theta_{\mathrm{E}}\right) + b$. The last column shows the Pearson correlation coefficient $r$ (see text for details).}
\label{table:correlation_theta_lcs}
\vspace{1.0em}
\centering
\begin{tabular}{lccccc}
\hline\hline
Cosmology & Einstein radius & \phantom{abcd} $\alpha$ \phantom{abcd} & \phantom{abcd}  $a$ \phantom{abcd} & \phantom{abcd} $b$ \phantom{abcd} &  \phantom{abcd} $r$ \phantom{abcd} \\
\hline

\multirow{4}{*}{WMAP7} & \multirow{2}{*}{$\theta_{\mathrm{E,med}}$} & $1.0$ & $2.44 \pm 0.03$ & $-5.68 \pm 0.03$ & $0.96$ \\
&  & $1.5$ & $2.31 \pm 0.02$ & $-5.35 \pm 0.03$ & $0.98$ \\
\cline{2-6}
& \multirow{2}{*}{$\theta_{\mathrm{E,eff}}$} & $1.0$ & $2.40 \pm 0.04$ & $-5.54 \pm 0.04$ & $0.95$ \\
&  & $1.5$ & $2.29 \pm 0.02$ & $-5.26 \pm 0.03$ & $0.97$ \\
\hline
\multirow{4}{*}{WMAP1} & \multirow{2}{*}{$\theta_{\mathrm{E,med}}$} & $1.0$ & $2.49 \pm 0.01$ & $-5.76 \pm 0.02$ & $0.96$ \\
&  & $1.5$ & $2.25 \pm 0.01$ & $-5.30 \pm 0.01$ & $0.98$ \\
\cline{2-6}
& \multirow{2}{*}{$\theta_{\mathrm{E,eff}}$} & $1.0$ & $2.46 \pm 0.02$ & $-5.63 \pm 0.02$ & $0.95$ \\
& & $1.5$ & $2.26 \pm 0.01$ & $-5.24 \pm 0.01$ & $0.98$ \\
\hline\hline
\end{tabular}
\end{table*}

The correlation qualitatively looks the same for all other settings and using effective Einstein radii $\theta_{\mathrm{E,eff}}$ instead. Hence we performed a least-squares fit for all data sets, assuming a linear relation of the form
\begin{equation}
\log\left(\sigma_{7.5}\right) = a\log\left(\theta_{\mathrm{E}}\right) + b \; .
\end{equation}
Additionally, we computed Pearson's correlation coefficient $r$. The results of this procedure are summarized in Tab. \ref{table:correlation_theta_lcs} and suggest the following conclusions. First, we note that the coefficients of the linear fits are not sensitive to the precise choice of the cosmological parameters. Furthermore, all Pearson correlation coefficients are close to unity, indicating that not only median, but also effective Einstein radii can be used to infer lensing cross sections in statistical studies. As may have been expected, the best-fittting linear relations between median and effective Einstein radii agree moderately.

\begin{figure}
	\centering
	\includegraphics[scale=0.95]{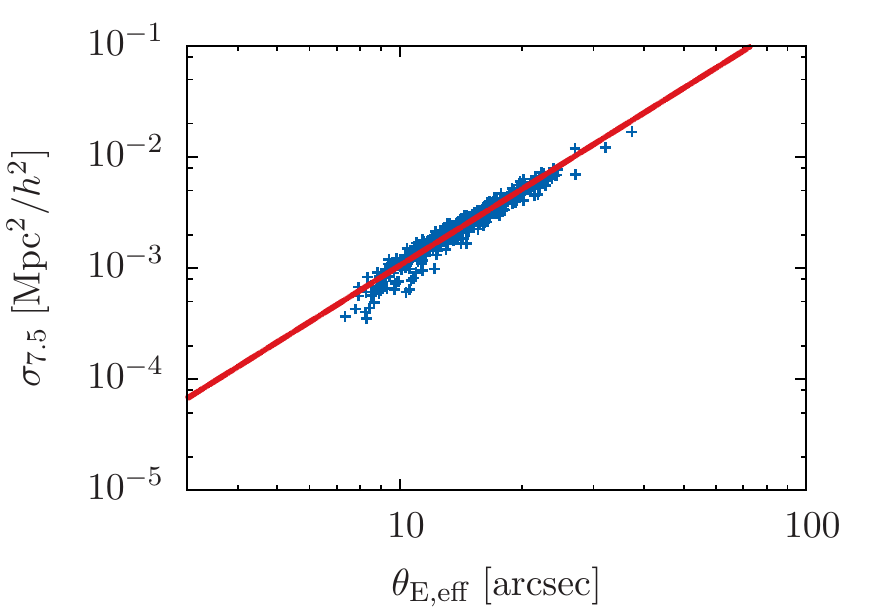}
	\caption[Correlation between effective Einstein radii and lensing cross sections for a cosmological population of triaxial dark matter haloes.]{Strong correlation between effective Einstein radii $\theta_{\mathrm{E,eff}}$ and lensing cross sections $\sigma_{7.5}$ for a cosmological population of triaxial dark matter haloes $\left(M > 5 \times 10^{14} M_{\odot} h^{-1}\right)$ at redshifts $z_{\rm l} > 0.5$, adopting a density profile with inner slope $\alpha = 1.5$. The population was drawn assuming a WMAP7 cosmology. The red solid line indicates the best linear fit relation between $\log(\sigma_{7.5})$ and $\log(\theta_{\mathrm{E,eff}})$.}
	\label{fig:theta_E_eff_vs_lcs_alpha15}
\end{figure}

Next, steeper density profiles produce slightly larger Einstein radii, while the lensing cross sections increase substantially \citep{2004PhDT.........1O}. This difference between steeper and flatter profiles manifests itself in systematically larger intercepts $b$ of the linear fits for steeper density profiles with inner slope $\alpha = 1.5$. Furthermore, it explains why the Pearson correlation coefficients indicate that the correlations are systematically stronger for $\alpha = 1.5$. The distribution of Einstein radii is shifted towards higher values and, as already mentioned, the overall scatter of the correlation decreases as the Einstein radii increase. Additionally, fewer Einstein radii fall into the region where the correlation breaks down. Both effects can be verified in Fig. \ref{fig:theta_E_eff_vs_lcs_alpha15}, which shows effective Einstein radii vs. lensing cross sections for the sample assuming a WMAP7 cosmology and adopting an inner slope $\alpha = 1.5$. Besides having different intercepts, we note that the slopes of the best linear fits are systematically flatter for $\alpha = 1.5$. We conclude that steeper density profiles lead to flatter correlations between Einstein radii and lensing cross sections. We confirmed this assumption with the following simple test. We generated another random sample of triaxial dark matter haloes and placed all lenses at one single redshift, namely $z_{\rm l} = 0.5$. In that case, the geometrical configuration of the lens system is such that arcs occur farther away from the halo centres where the logarithmic slope of the density profile is steeper. Consequently, we found a substantially flatter correlation for that specific sample.

\subsection{Comparison between numerical and analytical clusters}
\label{sec:comparison_correlation_numerical_sa}

\begin{figure}
	\centering
	\includegraphics[scale=0.9]{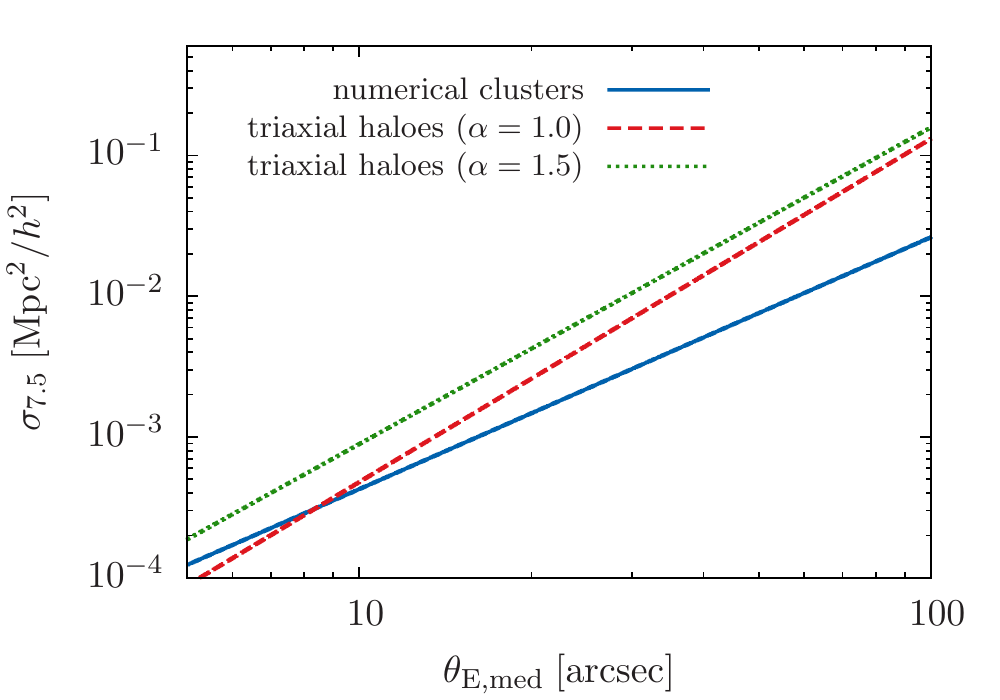}
	\caption[Best linear fits of the correlation between median Einstein radii and lensing cross sections for analytic and numerical cluster samples.]{Best linear fits of the correlation between median Einstein radii $\theta_{\mathrm{E,med}}$ and lensing cross sections $\sigma_{7.5}$ for numerical and analytic clusters. The blue solid line indicates the best linear fit for numerical clusters (cf. Eq. \eqref{eq:correlation_meneghetti}). The red dashed and the green dotted line indicate the best linear fits for cosmological populations of triaxial dark matter haloes (cf. Table \ref{table:correlation_theta_lcs}) with inner profile slopes $\alpha = 1.0$ and $\alpha = 1.5$, respectively, assuming a WMAP1 cosmology. The lensing cross sections of triaxial dark matter haloes are systematically larger in the range $\theta_{\mathrm{E,med}} \gtrsim 8 \arcsec$.}
	\label{fig:correlation_ns_vs_sa}
\end{figure}

We now compare our results to the findings of \citet{2011A&A...530A..17M} in detail. As indicated by Fig. \ref{fig:correlation_ns_vs_sa}, the lensing cross sections of triaxial haloes as a function of median Einstein radii are systematically larger than those of the numerical clusters throughout the entire range where the correlation between both quantities holds $\left(\theta_{\mathrm{E,med}} \gtrsim 8 \arcsec\right)$. This discrepancy even increases as the gravitational lenses become stronger. Since there is no obvious explanation for the above deviations, we briefly discuss the main differences between the two cluster samples. First, the \textsc{MareNostrum} simulation contains baryons, which are known to enhance the lensing cross sections of galaxy clusters \citep{2005A&A...442..405P, 2008ApJ...676..753W, 2008ApJ...687...22R, 2010MNRAS.406..434M}. Second, we describe dark matter haloes as ellipsoids with constant axis ratios, although numerical simulations indicate that these ratios slightly decrease towards the halo centres \citep{2002ApJ...574..538J}. Hence we slightly underestimate the ellipticity in the inner part of the density profile. Third, numerically simulated clusters naturally contain substructure and incorporate the effect of cluster mergers, whereas we only considered isolated dark matter haloes. All these previous aspects, i.e.\ asymmetry, substructure, and cluster mergers, are known to substantially increase the strong-lensing efficiency of galaxy clusters \citep{1995A&A...297....1B,2004MNRAS.349..476T, 2006A&A...447..419F, 2007MNRAS.381..171M}.

Given these differences, the lensing cross sections of numerical clusters should be significantly larger.  We speculate that the inclusion of substructure and cluster mergers is the main difference between the cluster samples. If this is indeed the case, Fig. \ref{fig:correlation_ns_vs_sa} suggests that these cluster properties become increasingly important for stronger gravitational lenses. Substructure leads to irregular critical curves with some extreme outliers. These, however, should be compensated for by the computation of \emph{median} Einstein radii. Hence, we expect that substructure, on average, increases the median Einstein radii only slightly, whereas lensing cross sections are likely significantly enhanced. Evidently, the comparison of the correlations in  Fig. \ref{fig:correlation_ns_vs_sa} indicates exactly the opposite trend, implying that substructure alone cannot explain the deviations. On the other hand, the evolution of lensing cross sections as a function of median Einstein radii during cluster mergers is far from obvious. In Section \ref{sec:mt_correlation}, we show that these events might indeed explain the discrepancy between the results of \citet{2011A&A...530A..17M} and our findings.

\section{The evolution of strong-lensing properties during cluster mergers}
\label{sec:evolution_slp_merger}

\subsection{Evolution of tangential critical curves}
\label{sec:mt_evolution_tangential_cc}

We now discuss the evolution of a tangential critical curve during a cluster merger by means of the following simplified toy model \citep[cf.][]{2004MNRAS.349..476T}. We considered two massive triaxial haloes merging at the fixed redshift $z_{\rm l} = 0.5$. The source plane was placed at $z_{\rm s} = 2.0$. The main halo $\left(M_{\mathrm{main}} = 1 \times 10^{15} \; M_{\odot} h^{-1} \right)$ was assumed to rest at the coordinate origin, while the second halo $\left( M_{\mathrm{sub}} = 2.5 \times 10^{14} \; M_{\odot} h^{-1} \right)$ was placed at an initial distance $d \sim 1\; \mathrm{Mpc} \; h^{-1}$ from the main halo. Since the precise properties (concentration, axis ratios, orientation) of both haloes do not matter for the following discussion, we simply state that their shape and orientation were chosen such that both lensing potentials are considerably elliptical. We simulated the merger by successively reducing the separation between both haloes in a series of discrete steps until their centres finally overlapped. At each step, we computed the resulting tangential critical curves.

\begin{figure*}
  \centering
  \subfloat{\includegraphics[trim=15 52 22 30,scale=0.9]{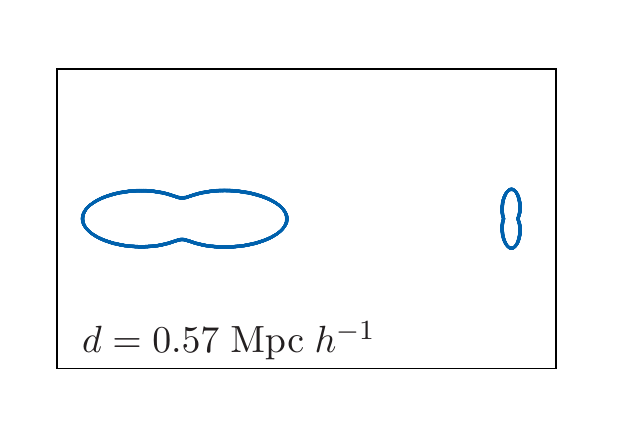}}
  \subfloat{\includegraphics[trim=15 52 22 30,scale=0.9]{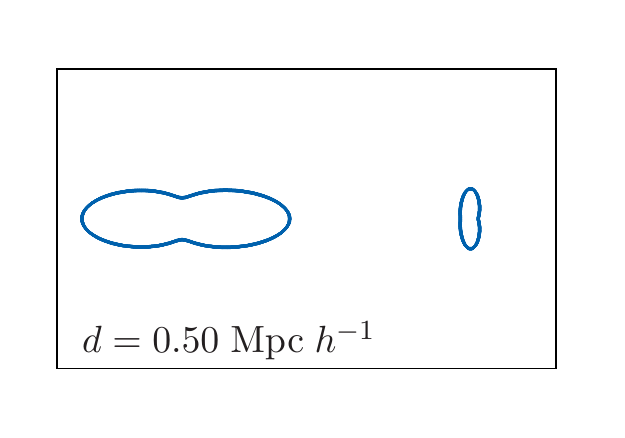}}
  \subfloat{\includegraphics[trim=15 52 50 30,scale=0.9]{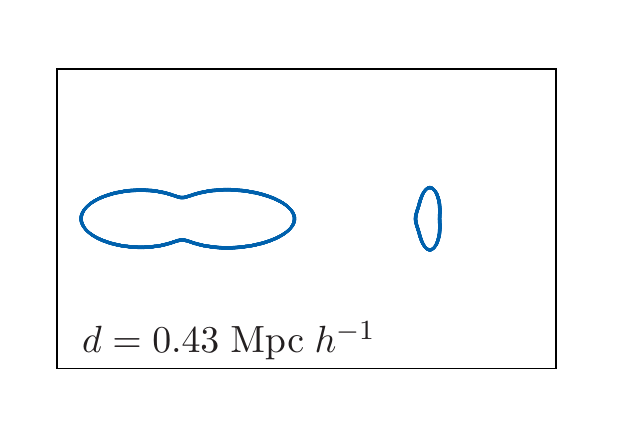}} \\
  \subfloat{\includegraphics[trim=15 52 22 0,scale=0.9]{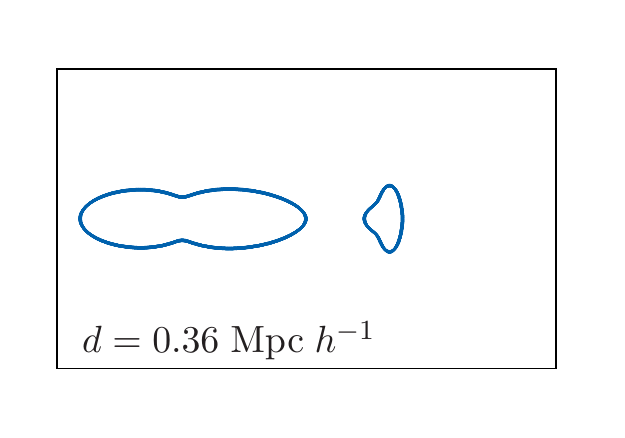}}
  \subfloat{\includegraphics[trim=15 52 22 0,scale=0.9]{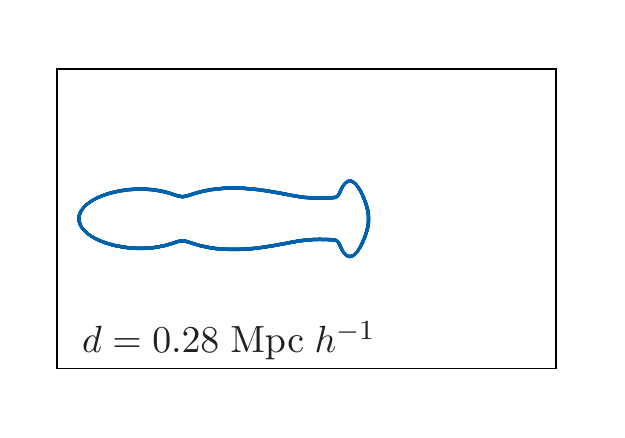}}
  \subfloat{\includegraphics[trim=15 52 50 0,scale=0.9]{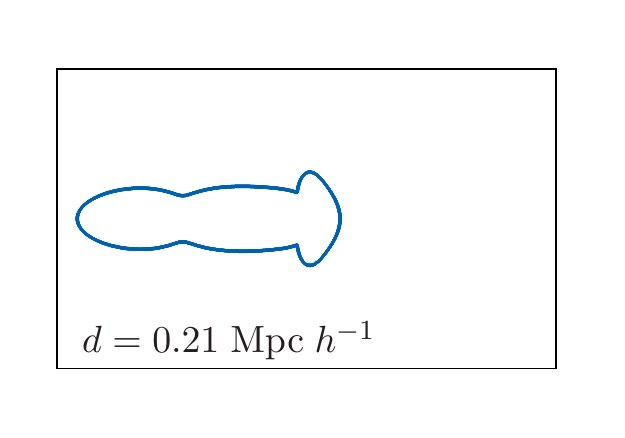}} \\
  \subfloat{\includegraphics[trim=15 15 22 0,scale=0.9]{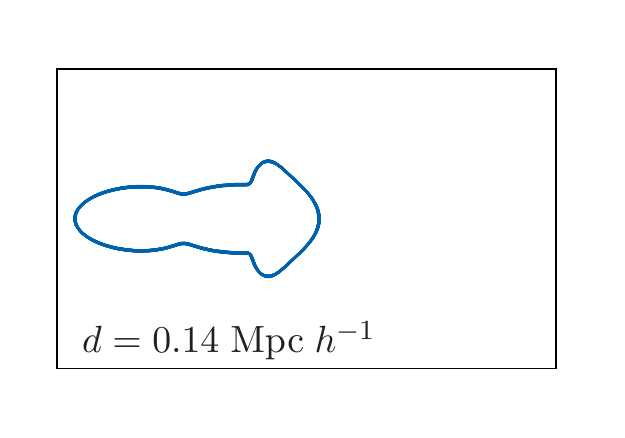}} 
  \subfloat{\includegraphics[trim=15 15 22 0,scale=0.9]{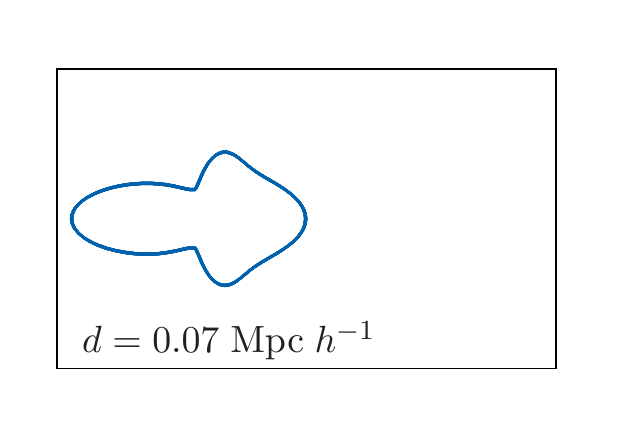}}
  \subfloat{\includegraphics[trim=15 15 50 0,scale=0.9]{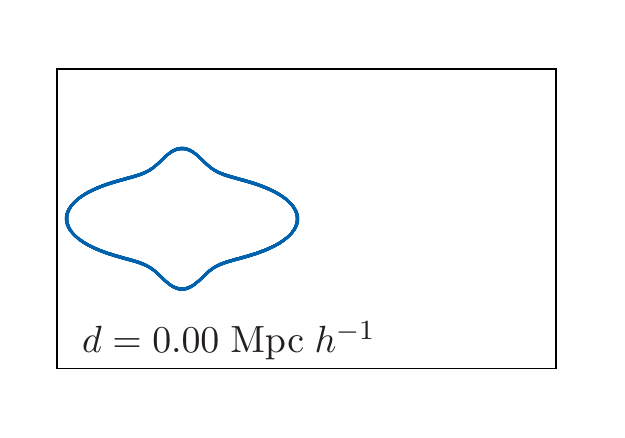}}
  \caption[Evolution of the tangential critical curves during a cluster merger.]{Evolution of the tangential critical curves during a merger of two massive $\left( M_{\mathrm{main}} = 1 \times 10^{15} \; M_{\odot} h^{-1}, M_{\mathrm{sub}} = 2.5 \times 10^{14} \; M_{\odot} h^{-1} \right)$ triaxial dark matter haloes. Starting with an initial separation of $d = 0.57 \; \mathrm{Mpc} \; h^{-1}$ in the top left panel, the distance between the two haloes is successively reduced until their centres finally overlap in the bottom right panel.}
  \label{fig:mt_evolution_cc}
\end{figure*}

Figure \ref{fig:mt_evolution_cc} illustrates the evolution of the tangential critical curves for a selection of intermediate steps of the simulated merger. As can be seen, we oriented the main halo such that the semi-major axis of its elliptical surface mass density profile was aligned with the direction of motion, whereas the elliptical projected mass profile of the second halo was rotated by 90 degrees. In the beginning (top left panel), the separation between the haloes is large so that both have their own isolated and almost unperturbed tangential critical curves. As soon as the haloes approach, the shear in the region between them grows and stretches both tangential critical curves along the direction of motion until they finally merge to build one large, highly elongated structure $\left(d \approx 0.32 \; \mathrm{Mpc} \; h^{-1} \right)$. Given the tight correlation between Einstein radii and lensing cross sections, we expect that these configurations are particularly efficient in producing thin arcs (see next section). As the smaller halo continues to approach the main halo, the highly elongated tangential critical curve typically starts to shrink along the direction of motion. Finally, however, the two mass profiles start to overlap significantly so that the convergence of the inner region grows substantially. As a consequence, the tangential critical curve again starts to expand almost isotropically.

This interpretation of the evolution of critical curves during cluster mergers was first given by \citet{2004MNRAS.349..476T}, and we refer the reader to their work for more details. It goes without saying that there are more sophisticated studies of cluster mergers that take virialization and other important physical processes properly into account. However, these models clearly go beyond the scope of our fast, semi-analytic approach to investigate the statistical strong-lensing properties of large cosmological populations of triaxial dark matter haloes.

\subsection{Evolution of Einstein radii and lensing cross sections}
\label{sec:relative_orientations}

\citet{2004MNRAS.349..476T} numerically simulated cluster mergers and found that these events are capable of boosting the lensing cross sections for giant arcs by about one order of magnitude. In a subsequent work, \citet{2006A&A...447..419F} showed that these events are not only important for the lensing cross sections of individual clusters, but also for the overall statistical strong-lensing efficiency of the cosmic cluster population. Motivated by these results, we performed a series of tests using the above toy model to study (1) whether we are able to reproduce the increase of the strong-lensing efficiency during mergers by modelling clusters with triaxial density profiles, (2) which parameters dominate the strength of this effect, and (3) whether the correlation between Einstein radii and lensing cross sections still holds during these events.

\begin{figure}
	\centering
	\includegraphics[scale=0.5]{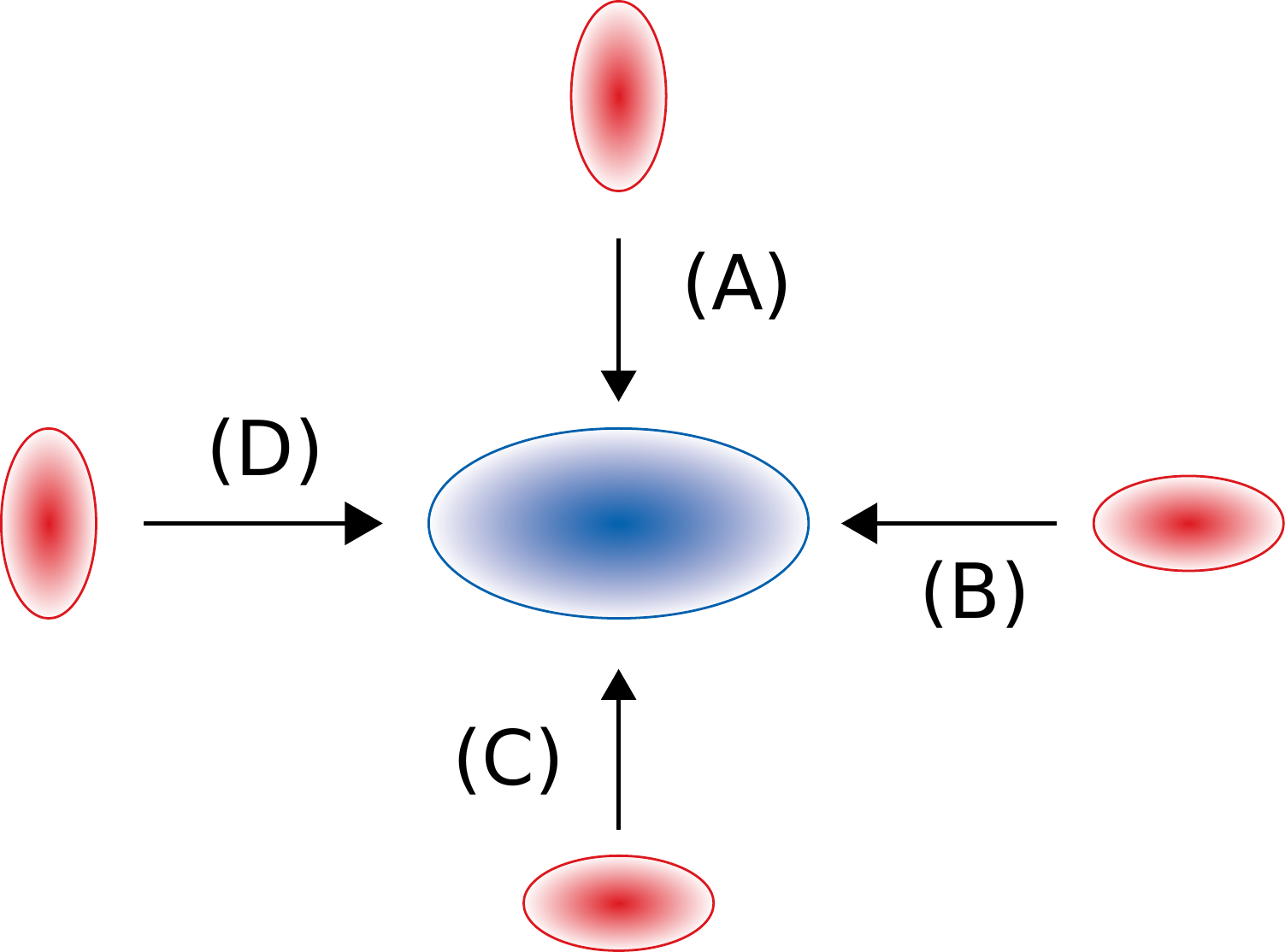}
	\caption[Sketch of four particular relative orientations of merging triaxial dark matter haloes.]{Schematic sketch of four particular orientations of merging triaxial dark matter haloes. The blue ellipsoid in the middle represents the projected mass density profile of the main halo. The red ellipsoids visualize the surface mass density profiles of the infalling subhaloes, which have different relative orientations with respect to the main halo. The black arrows indicate the directions of motion of the infalling structures.}
	\label{fig:merger_possible_orientations}
\end{figure}

Our most important results can be summarized as follows. Naturally, a notable enhancement of the strong-lensing efficiency can only be observed during major mergers (i.e. $M_{\mathrm{sub}}/M_{\mathrm{main}} \gg 0.05$). If the mass ratio $M_{\mathrm{sub}}/M_{\mathrm{main}}$ is too low, the elongation of the tangential critical curve enclosing the main halo is negligible. Conversely, the effect is particularly strong for mergers of clusters with comparable mass. Furthermore, the relative orientation of the ellipsoids plays a dominant role. Figure \ref{fig:merger_possible_orientations} sketches four particular relative orientations of merging triaxial dark matter haloes. The blue ellipsoid in the middle represents the projected mass density of the main halo, while the red ellipsoids display the surface mass densities of the infalling subhaloes. As discussed above, the enhancement of the strong-lensing efficiency can be observed shortly before the two initially separated tangential critical curves merge to form one large, extremely elongated structure. This effect is especially pronounced when the direction of motion and the semi-major axis of the main halo's surface mass density profile are aligned (mergers (B) and (D) in Figure \ref{fig:merger_possible_orientations}). The most extended tangential critical curves form in situation (B), where the semi-major axes of both projected mass profiles are aligned and additionally point along the direction of motion. Then, the elongation of the two separated tangential critical curves sets in relatively early and the diameter of the merged critical curve is particularly large. We emphasize that we can reproduce an increase of the lensing cross section by factors of $\sim 2$ for typical mass ratios $M_{\mathrm{sub}}/M_{\mathrm{main}} \sim 0.25$ if, and only if, this distinguished merger configuration is chosen and the surface mass density profiles of both haloes are appreciably elliptical (see below). In all other cases, the enhancement caused by mergers is notably smaller. Clearly, we cannot reproduce the order-of-magnitude increase found by \citet{2004MNRAS.349..476T}. Particularly in the least favourable configuration (C) there is almost no enhancement at all.

To quantify these general remarks, we exemplarily consider two specific cluster mergers in detail. In both cases, we simulated triaxial haloes of masses $M_{\mathrm{main}} = 1 \times 10^{15} \; M_{\odot} h^{-1}$ and $M_{\mathrm{sub}} = 2.5 \times 10^{14} \; M_{\odot} h^{-1}$, with axis ratios $a/c = 0.4$ and $b/c = 0.6$, mean concentration $\bar{c}_{\rm e} \approx 1.5$ (cf. Eq. \eqref{eq:concentration}), inner profile slope $\alpha = 1.0$ and projected along the $y'$-direction. 

\begin{figure*}
	\centering
	\subfloat{\includegraphics[scale=1.0]{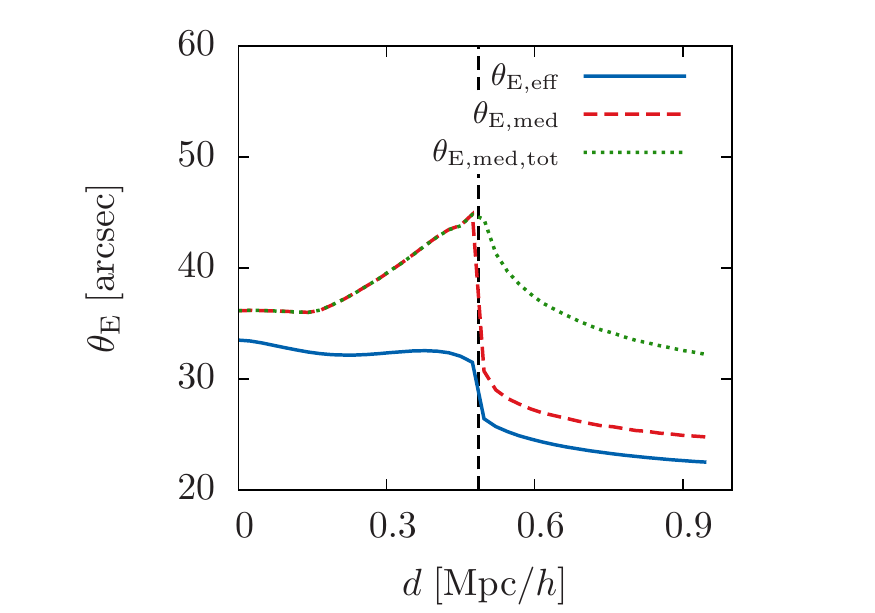}}
    \subfloat{\includegraphics[scale=1.0]{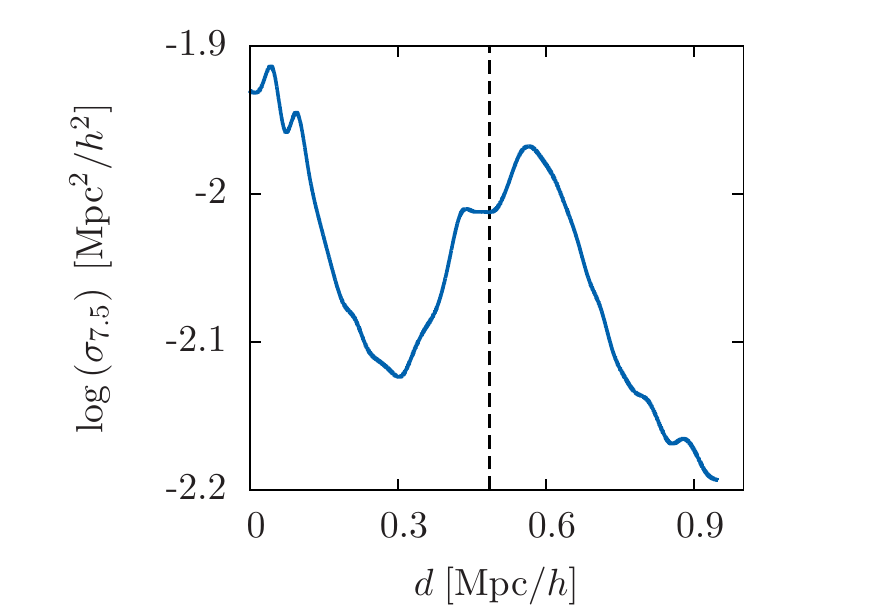}}
	\caption[Figure showing the evolution of Einstein radii and the lensing cross section during a cluster merger, assuming that the smaller halo approaches the main halo along the semi-major axis of its surface mass density profile.]{Figure showing the evolution of different definitions of Einstein radii (see text) and the lensing cross section during a simulated cluster merger, assuming configuration (B) in Figure \ref{fig:merger_possible_orientations}. All quantities are plotted as a function of the distance between both halo centres. The dashed vertical line indicates the moment when the tangential critical curves merge.}
	\label{fig:mt_slp_B}
\end{figure*}

We first consider a merger choosing orientation (D). Figure \ref{fig:mt_slp_B} shows the evolution of different definitions of Einstein radii and the lensing cross section as a function of the distance between the haloes' centres. In both diagrams, the dashed vertical line indicates the moment when the initially separated tangential critical curves merge. The blue solid and the red dashed lines in the left panel indicate the effective and median Einstein radii, respectively, computed for the tangential critical curve \emph{enclosing the main halo only}. These definitions take into account that both haloes initially have their own independent critical structures and are thus distinguishable systems from the lensing point of view. In contrast, the green dotted line indicates the median Einstein radius of all tangential critical points relative to the centre of the main halo. Here, "all" means that for the larger distances, where both clusters have their own critical curves, also the critical points enclosing the merging subhalo are taken into account for computing the median Einstein radius. This definition closely resembles that used by \citet{2011A&A...530A..17M}, who measured the distances of all tangential critical points with respect to the highest mass peak. We note that the effective Einstein radius seems to capture the isotropic expansion of the tangential critical curve as the density profiles of the two haloes start to overlap significantly and the convergence of the central regions grows. However, it does not indicate a substantial boost of the strong-lensing efficiency when the critical curves merge. This effect is better captured by median Einstein radii, which on the other hand do not reflect the isotropic growth of the tangential critical curve for short distances. The right panel of Fig. \ref{fig:mt_slp_B} shows the evolution of the total lensing cross section $\sigma_{7.5}$ of the two clusters. Obviously, the strong-lensing efficiency grows substantially shortly before the critical curves merge. For that specific orientation, we observe an increase of $\sim 170 \%$ in comparison to the initial value. After that peak, the lensing cross section passes through a minimum -- reflecting the fact that the elongation of the tangential critical curve decreases and the corresponding caustic becomes less cuspy -- before it finally increases by $\sim 200\%$ as a consequence of the overlapping mass profiles.

\begin{figure*}
	\centering
	\subfloat{\includegraphics[scale=1.0]{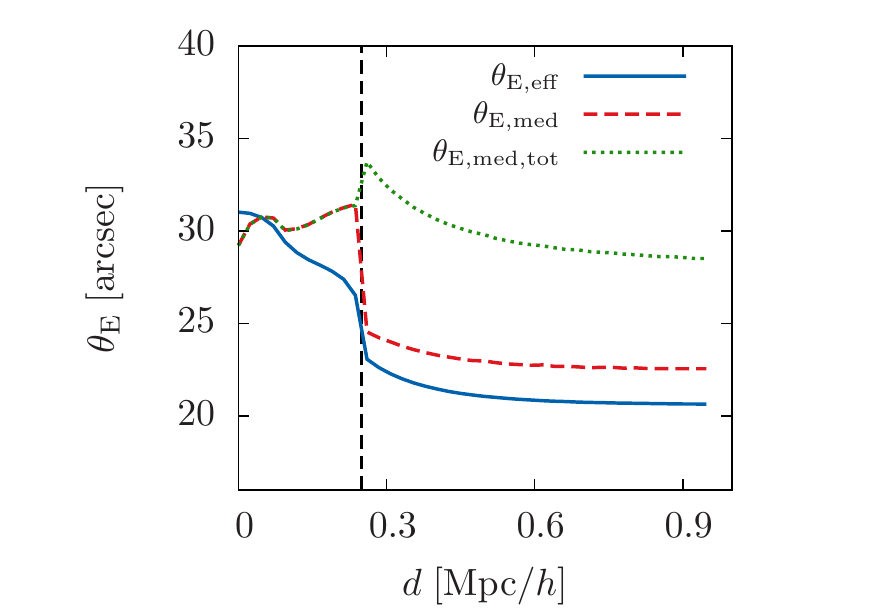}}
    \subfloat{\includegraphics[scale=1.0]{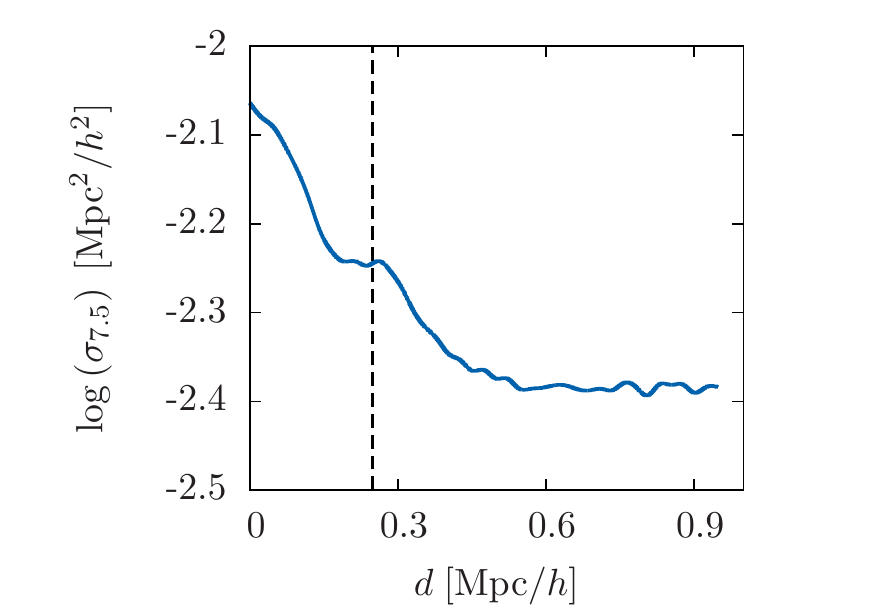}}
	\caption[Evolution of Einstein radii and the lensing cross section during a cluster merger, assuming that the smaller halo approaches the main halo along the semi-major axis of its surface mass density profile.]{Evolution of different definitions of Einstein radii (see text) and the lensing cross section during a simulated cluster merger, assuming configuration (A) of Figure \ref{fig:merger_possible_orientations}. All quantities are plotted as a function of the distance between the two halo centres. The dashed vertical line indicates the moment when the tangential critical curves merge.}
	\label{fig:mt_slp_a}
\end{figure*}

For comparison, Fig. \ref{fig:mt_slp_a} shows the same diagrams for orientation (A), which corresponds to a direction of motion perpendicular to the major axis of the main halo's surface mass density profile. Evidently, the evolution of the strong-lensing properties differs substantially from the previous merger. The elongation of the critical curve surrounding the central halo sets in later and the resulting Einstein radii are smaller. Moreover, the separation between the two clusters needs to be smaller for the critical curves to merge. The first peak of the lensing cross section is significantly smaller and coincides with the increase caused by the growing mass concentration in the central regions.

\subsection{Correlation between Einstein radii and lensing cross sections}
\label{sec:mt_correlation}

Our main motivation for studying the evolution of Einstein radii and lensing cross sections is to investigate whether or not the correlation between the two quantities still holds during mergers. If the relation is still valid, we can later simply compute all Einstein radii $\left( \theta_{\mathrm{E}} \gtrsim 8 \arcsec \right)$ on the sky and infer their lensing cross sections by means of the correlation instead of actually computing them with one of the discussed methods (cf. Section \ref{sec:lcs}). This approach would save much computing time. This final application may also justify why we plotted different definitions of Einstein radii in the left panels of Fig. \ref{fig:mt_slp_B} and \ref{fig:mt_slp_a}, since the first two of them only consider the tangential critical curve enclosing the main halo and are hence clearly more appropriate for our purpose.

The following analysis focuses only on the regions where both critical curves have already merged (i.e. left of the dashed vertical lines in Figs. \ref{fig:mt_slp_B} and \ref{fig:mt_slp_a}). First, we observe that none of the definitions of Einstein radii is able to preserve the tight correlation with lensing cross sections. In any case, there are regions where the Einstein radii decrease while the lensing cross sections increase, and vice versa. The correlation seems to break down completely if \emph{median} Einstein radii are used. In comparison, \emph{effective} Einstein radii trace the evolution of lensing cross sections better. However, by our definition of measuring only the area that is enclosed by the tangential critical curve surrounding the main halo, they do not capture the enhancement of the strong-lensing efficiency shortly before the tangential critical curves merge. We observe in passing that the definition of \emph{effective} Einstein radii is more natural for cluster mergers, because it does not require an arbitrarily chosen fiducial point.

\begin{figure*}
  \centering
  \subfloat{\includegraphics[trim=20 0 15 10,scale=1.1]{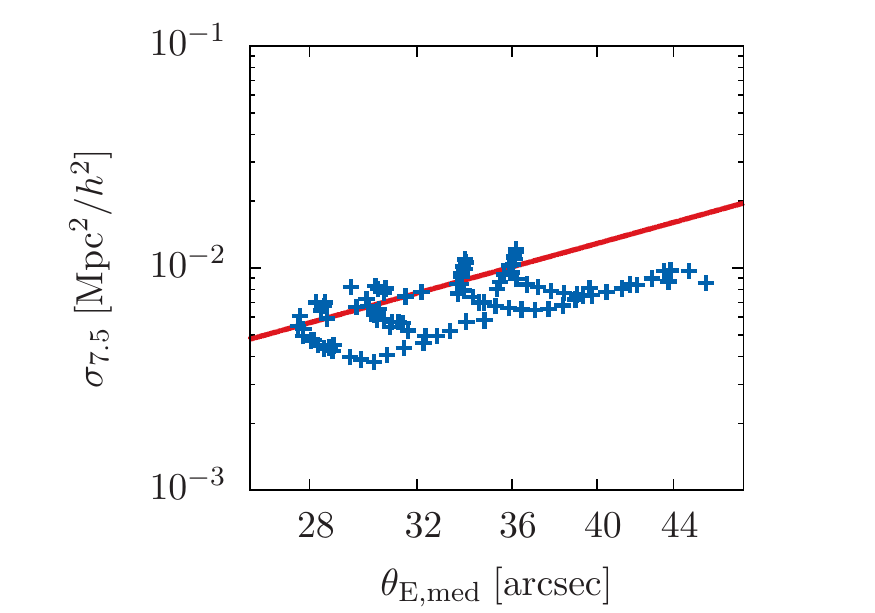}}
  \subfloat{\includegraphics[trim=15 0 20 10,scale=1.1]{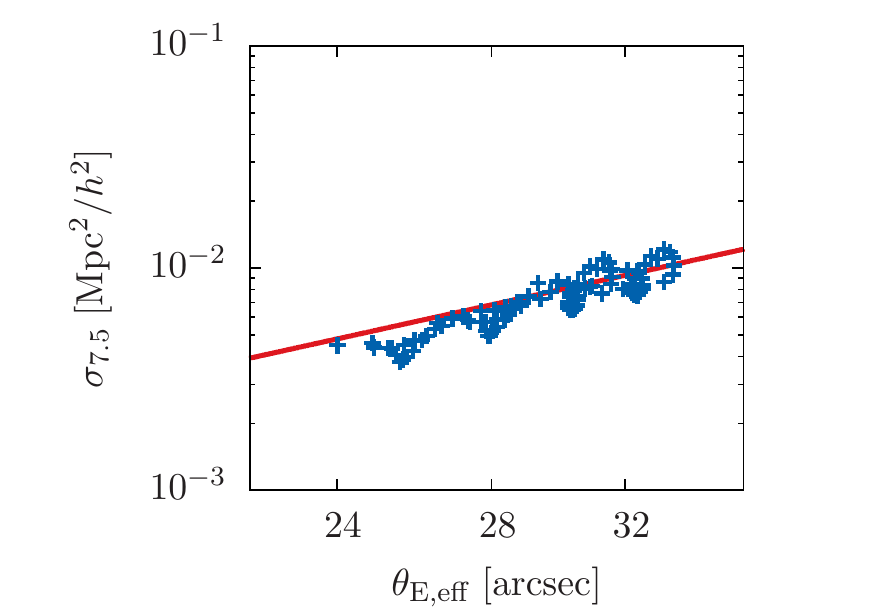}}
  \caption[Comparison of the correlation between Einstein radii and lensing cross sections for a sample of isolated triaxial dark matter haloes to the same correlation for a sample of merging triaxial dark matter haloes.]{Comparison of the correlation between Einstein radii and lensing cross sections for a sample of isolated triaxial dark matter haloes to the same correlation for a sample of merging triaxial dark matter haloes. The red solid line in the left (right) diagram indicates the best linear fit relation between median (effective) Einstein radii and lensing cross sections found for a sample of triaxial dark matter haloes at redshift $z_{\rm l} = 0.5$. The blue crosses represent selected pairs of Einstein radii and lensing cross sections during cluster mergers (see text for details).}
  \label{fig:mt_theta_lcs}
\end{figure*}

In Sect. \ref{sec:mt_evolution_tangential_cc}, we assumed a constant lens redshift of $z_{\rm l} = 0.5$ for our simplified models of cluster mergers. For reference, we therefore again generated a sample of 300 random triaxial haloes (mass range $5 \times 10^{14} - 2 \times 10^{15} \; M_{\odot} h^{-1}$, inner profile slope $\alpha = 1.0$) placed at $z_{\rm l} = 0.5$ and computed the best linear fits for the correlation between the two different definitions of Einstein radii and lensing cross sections. Figure \ref{fig:mt_theta_lcs} plots these fits, where the left and the right diagrams correspond to median and effective Einstein radii, respectively. In addition, we simulated all four merger configurations of Fig. \ref{fig:merger_possible_orientations} using the same dark matter haloes as in the previous section and computed their Einstein radii and lensing cross sections at each discrete simulation step. The blue crosses in the left and right diagrams indicate the results of all those discrete steps that correspond to distances with merged tangential critical curves. Evidently, for median Einstein radii, the pairs $\left(\theta_{\mathrm{E,med}},  \sigma_{7.5} \right)$ systematically lie below the correlation found for isolated clusters. This can easily be understood from Figs. \ref{fig:mt_slp_B} and \ref{fig:mt_slp_a}. Starting at $d = 0.0$, the median Einstein radii increase as a function of distance, whereas the lensing cross sections decrease, so there is indeed an anti-correlation. This observation provides a possible explanation for the discrepancy between the linear relation found by \citet{2011A&A...530A..17M} and our results (cf. Section \ref{sec:comparison_correlation_numerical_sa}). We expect that a certain fraction of Meneghetti and coworkers' numerically simulated clusters are actively merging so that their contribution flattens the results of the least-squares fit. If that line of reasoning is correct, we expect that the correlations found for numerical clusters predict smaller lensing cross sections as a function of Einstein radii. It would be an interesting task to verify this assumption by studying selected cluster samples in numerical simulations. The right-hand side diagram of Fig. \ref{fig:mt_theta_lcs} shows that the pairs of effective Einstein radii and lensing cross sections during cluster mergers agree significantly better with the corresponding linear fit found for isolated dark matter haloes. Hence we conclude that effective Einstein radii capture the evolution of strong-lensing properties during cluster mergers better and the correlation can still be used to approximate lensing cross sections in statistical lensing studies.

\section{Impact of cluster mergers on the statistics of the largest Einstein radii}
\label{sec:statistics_er}

\subsection{A semi-analytic approach to sample large cosmological halo populations that incorporates cluster mergers}
\label{sec:new_method_cluster_mergers}

While large cosmological $N$-body simulations are probably the most realistic framework for studying the impact of mergers on the statistical lensing properties of selected cluster samples, they are computationally costly. As an alternative, we now introduce a fast, semi-analytic method that allows us to study the statistical impact of mergers within a fraction of the computing time required for $N$-body simulations. We follow this approach because we aim to be able to repeat our calculations for large cosmic volumes, varying a multitude of boundary conditions (i.e. different properties of the triaxial density profile, cosmological parameters, mass functions, etc.), within a reasonably short time. This ability is important for our follow-up work, where we plan to employ methods of extreme value statistics to investigate the probability distribution of the largest Einstein radius expected within a $\Lambda$CDM cosmology \citep{2012arXiv1207.0801W}. We require our new approach to be fully self-consistent within the framework of the extended Press-Schechter theory \citep{1993MNRAS.262..627L}, which means that in particular the predicted merger rates and the Press-Schechter mass function \citep[cf.][]{1974ApJ...187..425P} must be reproduced accurately. This way, our results should be comparable with those derived from numerical simulations and real observations.

\begin{figure}
	\centering
	\includegraphics[scale=0.25]{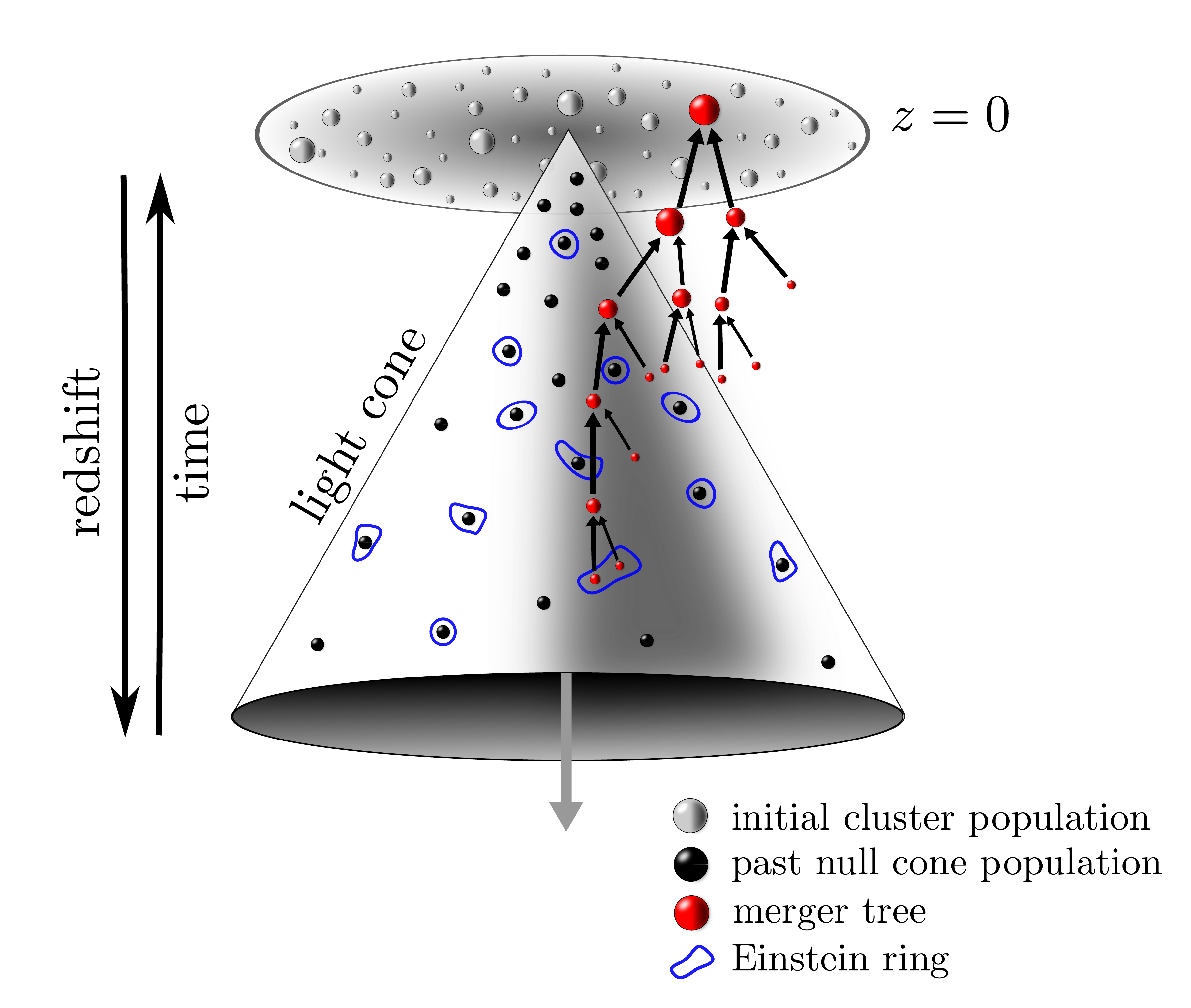}
	\caption[Sketch illustrating our new algorithm for generating realistic cosmological halo populations that incorporates cluster mergers.]{Sketch illustrating our new algorithm for generating realistic cosmological halo populations that incorporates clusters mergers. An initial halo population at redshift $z = 0$ is evolved backwards in time using extended Press-Schechter merger trees and is finally projected onto the observer's past null cone.}
	\label{fig:light_cone_sketch}
\end{figure}

Let us assume that we intend to analyse the strong-lensing properties of a representative all-sky realization of dark matter haloes in the mass and redshift range $\left[M_{\mathrm{min}}, M_{\mathrm{max}} \right]$ and $\left[z_{\mathrm{min}}, z_{\mathrm{max}} \right]$, respectively. To do this, we sequentially perform the following steps:

\begin{enumerate}

\item \emph{Initial halo population}: We use a common Monte Carlo approach (cf. \citet{2009MNRAS.392..930O} and Appendix \ref{app:drawing_cosmological_populations}) to populate the considered cosmic volume with an initial sample of dark matter haloes at the present time ($z=0$). We subdivide the mass range into logarithmically equidistant bins of typical size $\Delta \left( \log(M) \right) = 0.02$ and calculate the mean expected number $\bar{N}$ of haloes in each bin using the Press-Schechter mass function $\mathrm{d}n(M,z=0)/\mathrm{d}M$,
\begin{equation}
\bar{N} = \frac{\mathrm{d}n(M,z=0)}{\mathrm{d}M} \, \Delta M \times V(z_{\mathrm{min}}, z_{\mathrm{max}})   \;.
\end{equation}
Here, $V(z_{\mathrm{min}}, z_{\mathrm{max}})$ denotes the comoving volume of the spherical shell between the minimum and the maximum redshift (with respect to an observer at the coordinate origin). We then generate a random integer number $N$ from the Poisson distribution with mean $\bar{N}$ and sample $N$ haloes with logarithmically uniformly distributed masses in the corresponding bin. Since strongly lensing clusters are rare objects that are generally separated by large distances, we can neglect any kind of large scale structure correlation and simply uniformly distribute these $N$ sampled haloes in the considered cosmic volume. To this end, we assign each halo a random position on the sky, generate a uniformly distributed random number $x$ in the range $[0,1)$ and then compute the comoving distance $r$ with respect to the observer,
\begin{equation}
r = \left[ x r^3(z_{\mathrm{max}})+(1-x)r^3(z_{\mathrm{min}}) \right]^{1/3} \; ,
\end{equation}
where $r(z_{\mathrm{min}})$ and $r(z_{\mathrm{max}})$ denote the comoving distance to the minimum and maximum redshift, respectively. This formula takes the varying volume (as a function of radius) of spherical shells into account. Given the comoving radius $r$, we can easily infer each halo's redshift $z_{\mathrm{obs}}$ with respect to the observer.

\item \emph{Reverse time evolution}: We evolve the initial halo population backwards in time. More precisely, we adopt the efficient Monte Carlo approach proposed by \citet[][see 'method B']{2008MNRAS.389.1521Z} to simulate a representative merger tree of each individual halo up to its previously determined observation redshift $z_{\mathrm{obs}}$. Since \citet{2008MNRAS.389.1521Z} provide a nicely written step-by-step description of 'method B', we refer the reader to their work for details. Generally, merger tree algorithms generate representative formation histories by evolving the initial halo mass backwards in time taking discrete time steps (typical step size $\Delta z \approx 0.02$). At the first time step, the halo is split into smaller progenitors by means of the conditional mass function \citep{1993MNRAS.262..627L}. At the next time step, these new progenitors are again split into yet smaller progenitors using the same recipe, and so forth. Applying that scheme to all arising progenitors, this finally yields the full, tree-like formation history (i.e. the merger tree) of the initial halo. Given our predefined mass range, however, we discard all progenitors with masses below $M_{\mathrm{min}}$. We note that we verified that our implementation accurately reproduces the Press-Schechter mass function as well as the theoretically predicted merger rates at any look-back time.

\item \emph{Kinematics of merger trees}: To follow the trajectories of all arising progenitors in the considered cosmic volume, we additionally need to describe the kinematics of the merger trees. For that purpose, we adopt the following simplistic approach, which was first introduced by \citet{2007A&A...461...49F}. Each time a merger between two haloes of masses $M_1$ and $M_2$ occurs, we estimate its duration using the dynamical time scale
\begin{equation}
T_{\mathrm{dyn}} = \sqrt{\frac{\left(r_{\mathrm{vir},1} + r_{\mathrm{vir},2} \right)^3}{G\left( M_1 + M_2 \right)}} \; ,
\end{equation}
where $r_{\mathrm{vir},1}$ and $r_{\mathrm{vir},2}$ denote the virial radii of both haloes. These dynamical time scales are typically of the order of a several hundred Myr, which agrees well with merger durations measured in numerical simulations \citep[see][for instance]{2004MNRAS.349..476T}. Assuming a uniform linear motion, we compute the relative velocity of the two haloes,
\begin{equation}
v_{\mathrm{rel}} = \frac{r_{\mathrm{vir},1} + r_{\mathrm{vir},2}}{T_{\mathrm{dyn}}} \: ,
\end{equation}
and finally sample a random direction of motion. We apply this scheme to all mergers. This way, we can easily compute the spatial positions of all progenitors of the considered merger tree at the observation redshift $z_{\mathrm{obs}}$. We emphasize that our model of cluster mergers is clearly simplistic and neglects certain physical details. For instance, we implicitly assumed central cluster collisions although the centres of merging clusters can typically be separated by up to several hundred kpc \citep[see e.g.][for a semi-analytic approximation]{2000S..272....1S}. Finite impact parameters may certainly influence the detailed evolution of the strong-lensing properties during individual cluster mergers, and it would be interesting to study refinements of our simplified model in future works. However, we do not expect that these simplifications have a significant impact on the cumulative lensing efficiency of cosmological samples of merging clusters.

\end{enumerate}
Summarizing the above steps, we first populate the considered cosmic volume with an initial halo sample at redshift $z=0$, evolve that sample backwards in time using merger trees and a simplistic model to describe the kinematics, and finally project the resulting halo configurations onto the observer's past null cone. This procedure is illustrated in Fig. \ref{fig:light_cone_sketch}.

We conclude this section with a final remark on the specific choice of the mass function. Our algorithms are based on the Press-Schechter mass function \citep{1974ApJ...187..425P} throughout, although improved variants of the mass function with better accuracy have been proposed \citep[see][for instance]{2001MNRAS.321..372J, 2002MNRAS.329...61S, 2006ApJ...646..881W, 2008ApJ...688..709T}. This choice is motivated by the fact that the Press-Schechter mass function is based on the theory of spherical collapse and only within this framework exact analytic expressions for the conditional mass function -- which is the fundamental quantity for the computation of merger trees -- can be derived. In contrast, most improved variants are empirical fits to numerical simulations and hence there is no theoretical framework to derive the corresponding expressions for the conditional mass function. Since we aim to compare the statistics of Einstein radii of samples of isolated haloes to those of halo samples that incorporate cluster mergers in a fully self-consistent way, we are restricted to the mass functions based on the spherical (or ellipsoidal) collapse. The moderate deviations of the Press-Schechter mass function are negligible for our principal conclusions in this work. In contrast, if we were to perform a detailed comparison between our theoretical predictions and observations, the precise choice of the mass function would play an important role. This is mainly because the strongest gravitational lenses stem from the exponentially suppressed high-mass tail of the mass function, which is particularly sensitive to the exact calibration. Because the Press-Schechter mass function is known to underestimate the abundance of haloes in the high-mass tail, our approach also slightly underestimates the number of extreme lensing events. We refer the interested reader to our follow-up work \citep{2012arXiv1207.0801W}, where we provide a detailed discussion of the impact of different mass functions on the statistics of the single largest Einstein radius.

\subsection{Computation of Einstein radii}
\label{sec:computation_einstein_radii}

The previous step populates the observer's past null cone with (merging) dark matter haloes. We now describe an efficient method for computing the distribution of Einstein radii of that halo population.

First, we sort the masses in descending order and begin with the computation of the Einstein radius of the most massive halo $M_1$. To this end, we scan the region around that halo for neighbouring subhaloes $M_i$ whose distance $d$ is shorter than the sum of both virial radii, $d \leq \left(r_{\mathrm{vir},1} + r_{\mathrm{vir},i} \right)$. If any surrounding subhaloes satisfy this condition, they are taken into account for the computation of deflection angles. All other haloes that are farther away do not influence the strong-lensing properties of the considered massive halo $M_1$ and can safely be ignored. Next, we assign a random triaxial density profile and a random orientation with respect to the observer to each relevant halo. This enables us to determine the tangential critical curve that encloses the massive halo $M_1$. To do this, we place halo $M_1$ at the coordinate centre of a Cartesian grid whose side length is chosen to be sufficiently long to contain all relevant subhaloes. Instead of computing the deflection angle map of the complete field and determining the tangential critical curve afterwards, we simply need to detect the first tangential critical point left of the coordinate origin and employ a standard friend-of-friend algorithm to identify the entire tangential critical curve. This way, we only have to compute deflection angles in small stripes surrounding the critical curve, which renders the algorithm very fast. If we aimed to compute the lensing cross section of the considered configuration instead, we would not be able to use this efficient method but instead would have to calculate the complete deflection angle map. This important difference illustrates why the determination of Einstein radii is computationally far less expensive and why it is of great advantage to infer lensing cross sections from Einstein radii using the discussed correlation between both quantities. Finally, we detect all haloes which are enclosed by the computed tangential critical curve and mark them as haloes which do not have an independent Einstein radius. We repeat the above procedure for all remaining, unmarked haloes of the sorted mass list so that we end up with a complete catalogue of Einstein radii.

\subsection{Comparison of the statistics of Einstein radii with and without cluster mergers and inferred optical depths}

We can now quantify the statistical impact of cluster mergers on the strongest gravitational lenses $\left( \theta_{\mathrm{E,eff}} \ge 10 \arcsec \right)$ in our universe. To this end, we generated a first sample of single dark matter haloes using the conventional Monte Carlo approach (see Appendix \ref{app:drawing_cosmological_populations}) and a second sample using our newly developed technique to incorporate cluster mergers (see Section \ref{sec:new_method_cluster_mergers}). Since the largest observed Einstein radii, which may challenge the standard cosmological model, were all found at redshifts well below unity \citep[cf.][]{2008MNRAS.390.1647B, 2011MNRAS.410.1939Z}, we focused our analysis on the redshift range $z \in \left[ 0, 1\right]$. In anticipation of the following results, we note that the masses of all single haloes with effective Einstein radii above $10 \arcsec$ are larger than $10^{14} \, M_{\sun}/h$. Recalling that the mass ratio $M_{\mathrm{sub}}/M_{\mathrm{main}}$ of two merging clusters at least needs to exceed $5\%$ to notably boost the strong-lensing efficiency of the main halo, we could therefore safely ignore all haloes with masses below $M_{\mathrm{min}} = 5 \times 10^{12} \, M_{\sun}/h$. In particular, we pruned all branches (i.e. progenitors) of the merger trees whose masses dropped below that threshold. In this section, we modelled all haloes by means of triaxial density profiles with an inner slope $\alpha = 1$, adopted a WMAP7 cosmology and assumed that the source plane is placed at redshift $z_{\rm s} = 2.0$.

\begin{figure*}
  \centering
  \subfloat{\includegraphics[trim=0 0 0 0,scale=0.5]{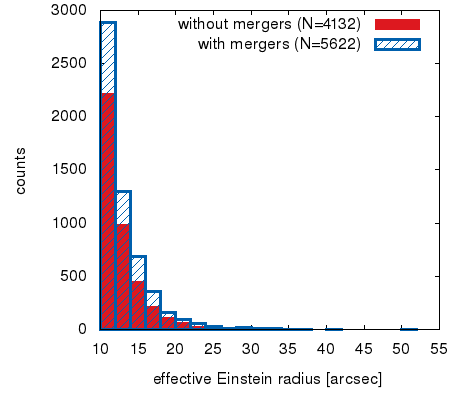}}
  \subfloat{\includegraphics[trim=0 0 0 0,scale=0.5]{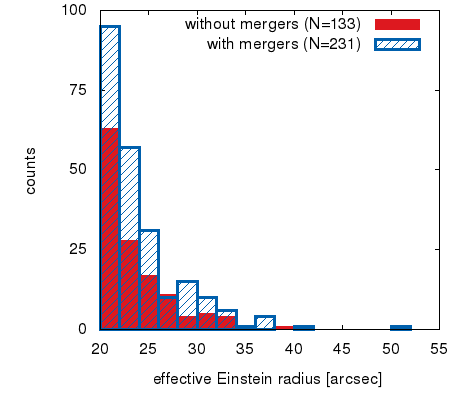}}
  \caption{Comparison of the distributions of the largest Einstein radii including (blue patterned histogram) and excluding (red solid histogram) cluster mergers. The left (right) plot shows the histograms of all Einstein radii larger than $10 \arcsec$ ($20 \arcsec$).}
  \label{fig:einstein_radii_comparison_histo}
\end{figure*}

Figure \ref{fig:einstein_radii_comparison_histo} and Tab. \ref{table:comparison_impact_merger} clearly demonstrate that cluster mergers do have a significant impact on the distribution of the largest Einstein radii. While we find $4132$ Einstein radii larger than $10 \arcsec$ in the single halo run, that number increases by $36\%$ up to $5622$ haloes if mergers are taken into account. The effect is even more significant if we only consider the number of systems with Einstein radii above $20 \arcsec$, in which case we find an increase of $74 \%$.  Another interesting aspect to be analysed is the frequency of cluster mergers. To do this, we classified observed systems as being actively merging if more than one halo is enclosed by the tangential critical curve. Clearly, this definition does not cover those cases where merging clusters are already close to each other, but still have their own, highly elongated critical curves.  Nevertheless, our wantedchoice should be sufficiently accurate for a qualitative estimate. We find that $35 \%$ of the systems with Einstein radii larger than $10 \arcsec$ are actively merging. This number increases to $55 \%$ if we only consider those clusters with Einstein radii larger than $20 \arcsec$. Given these results and the above number counts of Einstein radii, we can conclude that (1) cluster mergers are an important mechanism to increase the statistical lensing efficiency of cosmological halo populations and that (2) these events become increasingly dominant for particularly strong gravitational lenses.

\begin{figure*}
  \centering
  \subfloat{\includegraphics[trim=10 0 60 0,scale=0.85]{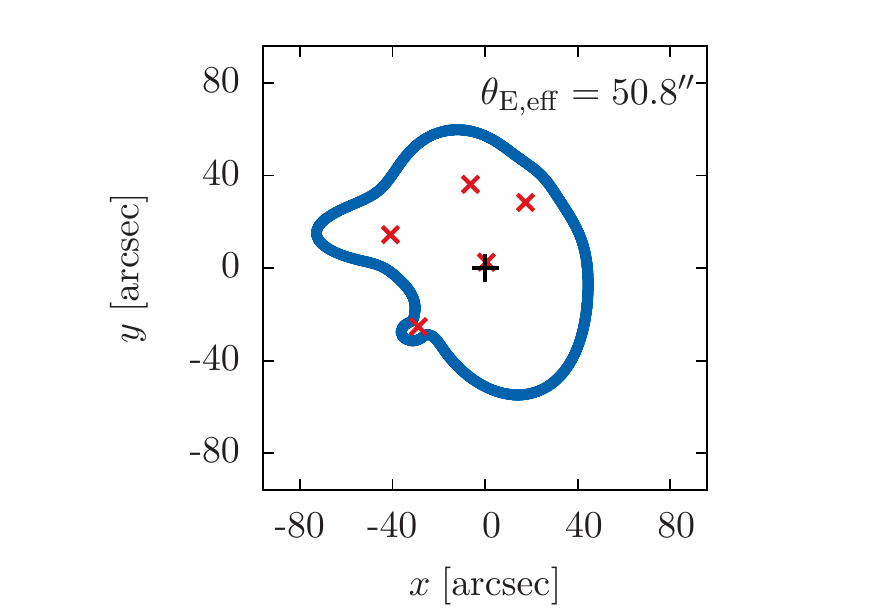}}
  \subfloat{\includegraphics[trim=10 0 60 0,scale=0.85]{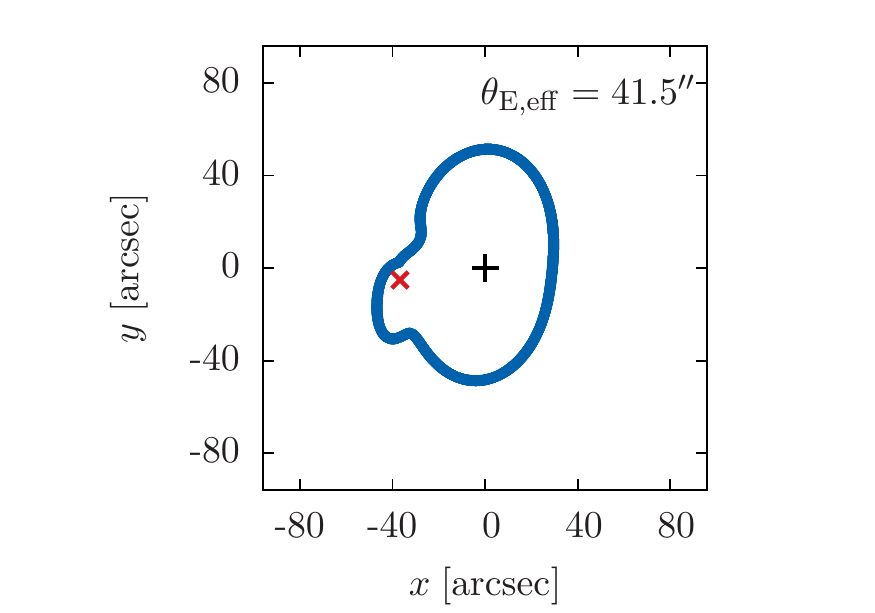}}
  \subfloat{\includegraphics[trim=10 0 60 0,scale=0.85]{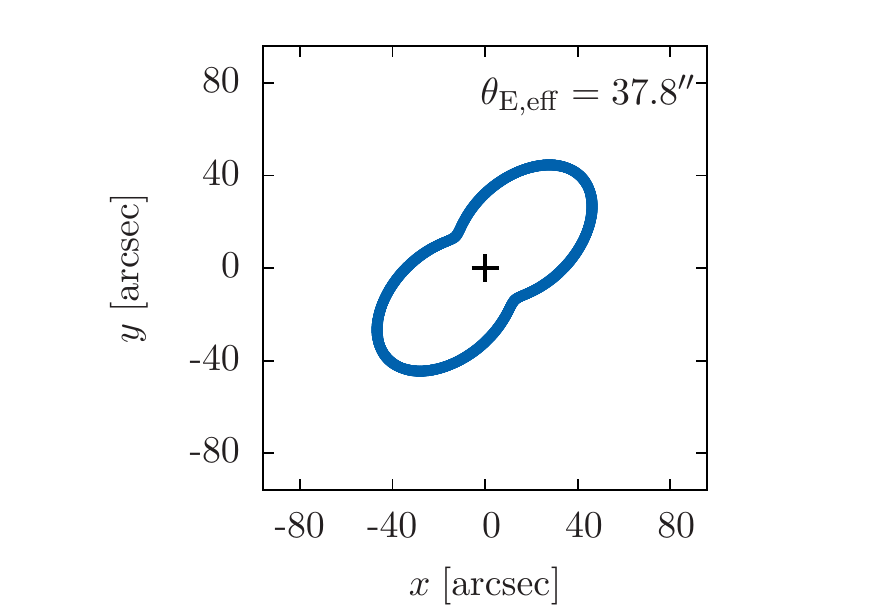}}
  \caption{Tangential critical curves of the systems with the largest (left diagram), second largest (middle diagram), and third largest (right diagram) Einstein radius in our simulation including cluster mergers. The black cross in the centre indicates the position of the main halo. The red crosses mark the positions of all subhaloes that are enclosed in the tangential critical curve.}
  \label{fig:einstein_radii_snapshots}
\end{figure*}

Additional evidence in favour of these conclusions is provided by considering the very largest lenses in detail. In the simulation including mergers, eight out of the ten largest Einstein radii stem from actively merging systems. Moreover, the largest Einstein radius in the single halo run has the size $\theta_{\mathrm{E,eff}} = 38.5 \arcsec$, while we find a notably larger maximum of $\theta_{\mathrm{E,eff}} = 50.8 \arcsec$ in the run with mergers. Needless to say, these maxima are subject to statistical fluctuations. Since the considered halo populations were randomly drawn, it may well be that for other realizations the single halo maximum is larger than the one found with merging haloes. However, we can safely conclude that cluster mergers need to be taken into account in semi-analytic, statistical studies that aim to challenge the standard cosmological model on the basis of the largest observed Einstein radii. The left panel of Fig. \ref{fig:einstein_radii_snapshots} shows the critical curve of the largest observed Einstein radius, which belongs to an extraordinary system at redshift $z = 0.43$. Starting with an initial mass of $M = 1.8 \times 10^{15} \, M_{\sun}/h$ at redshift $z = 0$, the halo was split into $20$ progenitors with masses above the minimum mass threshold $M_{\mathrm{min}}$ before it finally reached the observer's past null cone. Six of these 20 progenitors are enclosed in the critical curve shown. The most massive halo in the centre has a mass of $M = 1.1 \times 10^{15} \, M_{\sun}/h$. The masses of the five surrounding subhaloes lie in the range $6.7 \times 10^{12} - 1.5 \times 10^{14} \, M_{\sun}/h$. To further demonstrate the importance of mergers, we removed these subhaloes and computed the Einstein radius of the main halo alone, finding that it drops by $\sim 20 \arcsec$ to now only $\theta_{\mathrm{E,eff}} = 30.9 \arcsec$. Moreover, we can use that system to verify our choice of the minimum mass threshold $M_{\mathrm{min}} = 5.0 \times 10^{12} \, M_{\sun}/h$. The smallest enclosed subhalo has a mass of $M = 6.7 \times 10^{12} \, M_{\sun}/h$, which is close to $M_{\mathrm{min}}$. If we ignore that halo for the computation of the tangential critical curve, the Einstein radius drops only slightly by $\sim 1.4 \%$ to $\theta_{\mathrm{E,eff}} = 50.1 \arcsec$, confirming that our mass cut-off only leads to negligible errors. There are mainly two reasons for introducing such a mass cut-off in general: (1) we can discard a considerable amount of small-mass branches of the merger trees and (2) we can limit the number of (irrelevant) subhaloes that need to be taken into account for the computation of the critical curves. Both aspects substantially improve the performance of our algorithm.

\begin{figure}
  \centering
  \includegraphics[scale=0.5]{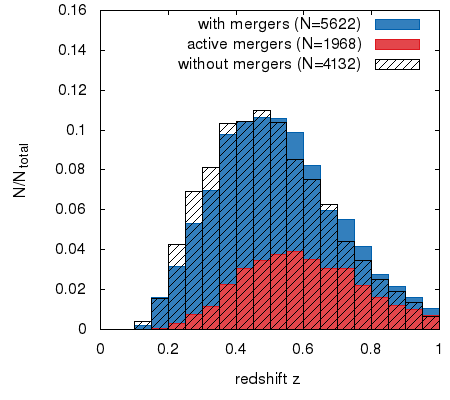}
  \caption{(Normalized) redshift distribution of the strong gravitational lenses with effective Einstein radii larger than $10 \arcsec$ including (blue solid histogram) and excluding (black patterned histogram) cluster mergers. In addition, the red stacks indicate the fraction of actively merging systems of all strong gravitational lenses represented by the blue solid histogram, i.e. including cluster mergers. Systems are classified as actively merging if more than one halo is enclosed by the tangential critical curve (see text).}
  \label{fig:redshift_distribution_sl_histo}
\end{figure}

The framework of hierarchical structure formation suggests that relatively young clusters at intermediate and high redshifts ($z \gtrsim 0.5$) are dynamically more active than older systems at low redshifts. Therefore, it is to be expected that mergers mainly increase the number of strong lenses in the upper redshift range of our simulation volume. Although Fig.~\ref{fig:redshift_distribution_sl_histo} confirms this expectation, the net shift of the overall population of strong-lensing clusters to higher redshifts due to mergers is only moderate. Instead, the redshift distribution of the strongest lenses is predominantly determined by the chosen lensing geometry with sources at redshift $z_{\mathrm{s}} = 2.0$.

Finally, we used the correlation between Einstein radii and lensing cross sections to infer the impact of cluster mergers on the statistics of giant gravitational arcs from the distributions of Einstein radii just computed. Since we concentrated on the strongest gravitational lenses only, we are well inside the regime where the correlation between both quantities is particularly tight and hence our following estimate should be reasonably accurate. First, we randomly picked $\sim 400$ haloes of our catalogue with Einstein radii above $10 \arcsec$ and computed their lensing cross sections $\sigma_{7.5}$. We then used these results to calibrate the best linear fit relation for the correlation:
\begin{equation}
\label{eq:theta_e_sigma_corr_final}
\log (\sigma_{7.5}) = (2.12 \pm 0.06) \log(\theta_{\mathrm{E,eff}}) - (5.24 \pm 0.07) \; .
\end{equation}
The cumulative lensing efficiency of cluster samples is usually characterized by their \emph{optical depth $\tau$} for giant gravitational arcs, which is given by the sum of the individual lensing cross sections $\sigma_{7.5, i}$ divided by the size of the entire source sphere,
\begin{equation}
\label{eq:optical_depth}
\tau \equiv \left(\sum\limits_{i} \sigma_{7.5,i}\right) \times \left(4 \pi D_s^2\right)^{-1} \; ,
\end{equation}
where $D_s$ denotes the angular diameter distance to the source plane. Assuming that the individual lensing cross sections do not overlap in the source plane, the optical depth corresponds to the probability that an arbitrarily placed source with the specified characteristics produces a gravitational arc with a length-to-width ratio higher than $7.5$. We find that cluster mergers increase the optical depth of all haloes with Einstein radii above $10 \arcsec$ by roughly $45 \%$. Furthermore, as was to be expected, the impact on those systems with Einstein radii above $20 \arcsec$ is even stronger: we find that mergers statistically increase the number of giant arcs produced by these particularly strong lenses by roughly $85 \%$.

\begin{table}
\caption{Comparison of the strong-lensing statistics of two representative samples of dark matter haloes ($M > 5 \times 10^{12} \, M_{\sun}/h, \, z \in [0, 1]$) including and excluding cluster mergers, adopting a WMAP7 cosmology and a constant source redshift $z_{\mathrm{s}} = 2.0$. We show the maximum effective Einstein radius $\mathrm{max}(\theta_{\mathrm{E,eff}})$ including the observation redshift $z_{\mathrm{obs}}$ of the corresponding lens system, the number of Einstein radii above a certain threshold $X$, $N\left( \theta_{\mathrm{E,eff}} \ge X \right)$, and the optical depth (for gravitational arcs, see Eq. \eqref{eq:optical_depth}) of all haloes with Einstein radii above the threshold $X$, $\tau (\theta_{\mathrm{E,eff}} > X)$.}
\label{table:comparison_impact_merger}
\vspace{1.0em}
\centering
\begin{tabular}{c|c|c}
 & single haloes & merging haloes \\ 
\hline
\hline
$\mathrm{max}(\theta_{\mathrm{E,eff}})$ & $38.5 \arcsec \left(z_{\mathrm{obs}} = 0.21\right)$ & $50.8 \arcsec \left( z_{\mathrm{obs}} = 0.43 \right)$ \\ 
\hline
$N \left( \theta_{\mathrm{E,eff}} \ge 10 \arcsec \right)$ & $4132$ & $5622 \; (+36\%)$ \\
\hline
$N \left( \theta_{\mathrm{E,eff}} \ge 20 \arcsec \right)$ & $133$ & $231 \; (+74\%)$ \\
\hline
$\tau (\theta_{\mathrm{E,eff}} > 10 \arcsec)$ & $2.9 \times 10^{-7}$ & $4.2 \times 10^{-7} \; (+45 \%)$ \\ 
\hline 
$\tau (\theta_{\mathrm{E,eff}} > 20 \arcsec)$ & $3.3 \times 10^{-8}$ & $6.1 \times 10^{-8} \; (+85\%)$ \\ 
\hline 
\end{tabular} 
\end{table}

\section{Conclusions}
\label{sec:conclusions}

Our main results can be briefly summarized as follows. As a first preparatory step, we carried out a detailed comparison between three alternative methods to compute lensing cross sections. We showed that the fast, semi-analytic method originally proposed by \citet{2006A&A...447..419F} systematically deviates from ray-tracing algorithms due to a differing definition of the length-to-width ratio of lensed images. However, this discrepancy can easily be resolved by a trivial correction, after which the semi-analytic method yields equally reliable results as the well-probed ray-tracing codes. Hence, our findings confirm that the semi-analytic method represents an extremely valuable tool for computationally demanding problems. Moreover, our analysis clearly indicates that the precise definition of the length-to-width ratio of lensed images has a significant impact on the final lensing cross section. Thus, special care must be taken to adopt consistent definitions, in particular when comparing theoretical predictions with real observations in the context of the arc statistics problem.

Secondly, we investigated the correlation between Einstein radii and lensing cross sections that was found recently by \citet{2011A&A...530A..17M} for a selected sample of numerical clusters. Considering single haloes only, we showed that we can reproduce the tight correlation, using both \emph{median} and \emph{effective} Einstein radii to measure the size of tangential critical curves. However, modelling cluster mergers by means of simplified toy models, we demonstrated that the strict correlation between \emph{median} Einstein radii and lensing cross sections breaks down during these events. In comparison, \emph{effective} Einstein radii capture the evolution of the lensing cross sections notably better. Furthermore, we found that the relative orientation of two merging triaxial haloes plays a significant role for the evolution of the strong-lensing properties. While we were able to reproduce an enhancement of the strong-lensing efficiency by factors of a few choosing the most beneficial relative orientation, we showed that this effect completely vanishes when averaging other alignments.

The correlation between effective Einstein radii and lensing cross sections is tight enough to infer sufficiently accurate estimates of optical depths for giant arcs from cosmological distributions of Einstein radii. This approach is computationally far less demanding than explicitly calculating individual lensing cross sections, since the computation of Einstein radii can be implemented in a particularly efficient way. We therefore developed a new, semi-analytic algorithm for computing cosmological distributions of Einstein radii that incorporates the impact of cluster mergers. Using this new technique, we were able to compare the statistical strong-lensing properties of one sample of single (isolated) dark matter haloes and a second sample of dark matter haloes that includes cluster mergers. Assuming a constant source redshift of $z_{\mathrm{s}} = 2.0$, we found that cluster mergers increase the theoretically expected number of Einstein radii above $10 \arcsec$ ($20 \arcsec$) by roughly $36 \%$ ($74 \%$), indicating that these events provide a highly efficient mechanism to enhance the lensing efficiency of particularly strong gravitational lenses. These results clearly show that semi-analytic studies need to take cluster mergers into account if they aim to question the standard cosmological model by comparing the largest observed Einstein radii to theoretical expectations. Furthermore, we estimated that the optical depth for giant gravitational arcs of those haloes with Einstein radii above $10 \arcsec$ ($20 \arcsec$) increases by roughly $45 \%$ ($85 \%$), which highlights the importance of cluster mergers for the statistics of giant gravitational arcs. Still, we were unable to reproduce the doubling of the optical depth found by \citet{2006A&A...447..419F}. However, this discrepancy can be attributed to the following important differences between both studies:

\begin{itemize}

\item Instead of adopting triaxial density profiles with varying properties (i.e. concentration and axis ratios), \citet{2006A&A...447..419F} modelled clusters as spherically symmetric density profiles and elliptically distorted their lensing potentials by a constant amount. Moreover, these authors used an alternative concentration-mass relation. Consequently, their lens model produces surface mass densities with different properties, which are decisive for the strength of the boost of the strong-lensing efficiency during cluster mergers.

\item Analyses of cluster mergers in numerical simulations reveal that (1) the major axes of infalling substructures are intrinsically aligned with the major axis of the main halo due to its tidal field and (2) subhaloes preferentially approach the main halo along its major axis \citep{2005ApJ...629L...5L, 2006MNRAS.370.1422A, 2009ApJ...706..747Z}. Motivated by these findings, \citet{2006A&A...447..419F} throughout modelled mergers by perfectly aligning the major axes of both haloes and assuming that the subhalo always approaches the main halo exactly along its major axis. As shown in Section \ref{sec:relative_orientations}, these orientations are most favourable and produce the strongest possible enhancement of the strong-lensing efficiency during cluster mergers. On the other hand, this assumption clearly is an idealization that might lead to an overestimation of the net effect of mergers. Since there is no reliable theoretical framework that allows us to predict those alignments realistically, we chose the fairly conservative approach to sample the relative orientations of merging haloes randomly, without taking their axis correlations and preferred directions of motion into account. Hence, we expect that our results slightly underestimate the net effect of mergers.

\item While we only computed the optical depth of particularly strong lenses with Einstein radii above $10 \arcsec$, \citet{2006A&A...447..419F} determined the optical depth of all clusters at redshifts $z > 0.5$. Hence, their computation additionally incorporates the following important effect. Cluster mergers tend not only to increase the lensing cross section of the strongest gravitational lenses, but are also capable of transforming clusters to strong gravitational lenses that would otherwise be subcritical. More precisely, it may well be that two relatively small clusters with vanishing individual lensing cross sections produce a nonvanishing, joint lensing cross section while merging. Consequently, mergers are an effective mechanism to substantially increase the optical depth of clusters in the lower mass range. Given the steepness of the mass function, these objects dominate the total optical depth and thus their contribution is likely significant for the doubling of the optical depth found by \citet{2006A&A...447..419F}.

\end{itemize}
Given these remarks, we conclude that our findings do not disagree with the previous results of \citet{2006A&A...447..419F}.

In this work, we mainly focused on the development of new techniques for incorporating the impact of cluster mergers into semi-analytic, statistical strong-lensing studies. We did not conduct a detailed comparison between theory and observations, in particular with regard to the problem of too large Einstein radii. Such analyses will be performed in our follow-up work, where we plan to apply methods of extreme value statistics to investigate whether or not cluster mergers are required (and sufficient) to explain the largest observed Einstein radii within the framework of the standard cosmological model \citep{2012arXiv1207.0801W}.

\begin{acknowledgements}

We are grateful to Massimo Meneghetti and Ewald Puchwein for kindly providing us their ray-tracing implementations for computing lensing cross sections, and we also acknowledge many stimulating discussions. We would also like to thank Francesco Pace for helpful discussions and his tireless translations from ancient \textsc{Fortran} scripts. Finally, we would like to thank the anonymous referee for his/her constructive und clarifying comments. MR's work
was supported in part by contract research 'Internationale Spitzenforschung II-1' of the Baden-W\"urttemberg Stiftung. JCW acknowledges financial contributions from contracts ASI- INAF I/023/05/0, ASI-INAF I/088/06/0, ASI I/016/07/0 COFIS, ASI Euclid-DUNE I/064/08/0, ASI-Uni Bologna-Astronomy Dept. Euclid-NIS I/039/10/0 and PRIN MIUR 2008 ‘Dark energy and cosmology with large galaxy surveys’. CF acknowledges support from the University of Florida through the Theoretical Astrophysics Fellowship. MB is supported in part by the Transregional Collaborative Research Centre TR 33 ``The Dark Universe'' of the German Science Foundation.

\end{acknowledgements}

\bibliographystyle{aa}
\bibliography{merger_strong_lensing}

\onecolumn

\begin{appendix}

\section{Gravitational lensing by triaxial dark matter haloes with generalized coordinate transformation}
\label{app:coordinate_transformation}

As mentioned in Section \ref{sec:lensing_tdmh}, we need to introduce a more general rotation matrix than \citet{2003ApJ...599....7O} to describe the coordinate transformation between the principal axis frame of triaxial haloes and the rest frame of the observer. Consequently, some algebraic expressions in the derivation of the lensing properties of triaxial haloes are more complicated. Adopting the same notation as \citet{2003ApJ...599....7O}, the corresponding expressions are given by

\begin{align}
\label{eq:f}
f &= \cos^2\theta + \sin^2\theta \left( \frac{c^2}{a^2} \sin^2\phi + \frac{c^2}{b^2} \cos^2\phi \right) \; , \\
\notag \\
g &= \sin\theta \left\{\sin \psi \cos \theta \left[ \cos 2\phi \left( \frac{c^2}{a^2} - \frac{c^2}{b^2} \right)  + 2 - \frac{c^2}{a^2} - \frac{c^2}{b^2} \right] + \cos \psi \sin 2\phi \left( \frac{c^2}{a^2} - \frac{c^2}{b^2} \right) \right\} \, x \\ 
 & \phantom{=} + \sin\theta \left\{ -\cos \psi \cos \theta \left[ \cos 2 \phi \left( \frac{c^2}{a^2} - \frac{c^2}{b^2}  \right) +2 - \frac{c^2}{a^2} - \frac{c^2}{b^2} \right]
 + \sin\psi\sin 2\phi \left( \frac{c^2}{a^2} - \frac{c^2}{b^2} \right) \right\} \, y \; , \notag \\
\notag \\
h &= \left[ \sin^2\psi \sin^2\theta + \frac{c^2}{a^2} \left( \cos \psi \cos \phi - \sin \psi \cos \theta \sin \phi \right)^2 + \frac{c^2}{b^2} \left( \cos \psi \sin \phi + \sin \psi \cos \theta \cos \phi \right)^2 \right] \, x^2 \\
& \phantom{=} + \Bigg\{ - \sin 2 \psi \sin^2 \theta + 2 \frac{c^2}{a^2} \left[ \left( \cos \psi \cos \theta \sin \phi + \sin\psi \cos \phi \right)  \left( \cos \psi \cos \phi - \sin \psi \cos \theta \sin \phi \right) \right] \notag \\
& \qquad \qquad - 2 \frac{c^2}{b^2} \left[ \left( \sin\psi\cos\theta\cos\phi + \cos \psi \sin\phi \right) \left( \cos\psi\cos\theta\cos\phi - \sin\psi\sin\phi \right) \right] \Bigg\} \, xy \notag \\
& \phantom{=} + \Bigg[ \cos^2 \psi \sin^2 \theta + \frac{c^2}{a^2} \left( \cos \psi \cos \theta \sin \phi + \sin \psi \cos \phi \right)^2 + \frac{c^2}{b^2} \left( \cos\psi \cos\theta\cos\phi - \sin\psi\sin\phi \right)^2 \Bigg] \, y^2 \; . \notag \\
\notag \\ 
A & \equiv \Bigg[ \sin^2\theta \left( \frac{c^2}{a^2}\sin^2\phi + \frac{c^2}{b^2}\cos^2\phi \right) + \cos^2\theta \Bigg] \times \Bigg[ \frac{c^2}{a^2} \left( \cos\psi\cos\phi - \sin\psi \cos\theta \sin\phi \right)^2 \\  
 & \qquad \qquad \qquad \qquad \qquad \qquad \qquad \qquad \qquad \qquad + \frac{c^2}{b^2} \left( \sin\psi \cos\theta \cos\phi + \cos\psi \sin\phi \right)^2 + \sin^2\psi \sin^2\theta \Bigg] \notag \\ 
 & \phantom{\equiv} - \frac{\sin^2\theta}{4} \Bigg\{ \sin\psi \cos\theta \left[ \cos 2\phi \left( \frac{c^2}{a^2} - \frac{c^2}{b^2} \right) + 2 - \frac{c^2}{a^2} - \frac{c^2}{b^2} \right] + \cos\psi \sin 2\phi \left( \frac{c^2}{a^2} - \frac{c^2}{b^2} \right) \Bigg\}^2 \; , \notag \\
\notag \\
B & \equiv \quad \left[ \sin^2\theta \left(\frac{c^2}{a^2} \sin^2\phi + \frac{c^2}{b^2}\cos^2\phi \right) + \cos^2\theta \right] \\ 
& \qquad \qquad \times \Bigg[ 2 \frac{c^2}{a^2} \left( \cos\psi\cos\theta\sin\phi + \sin\psi\cos\phi\right) \left( \cos\psi\cos\phi - \sin\psi\cos\theta\sin\phi \right) \notag \\
 & \qquad \qquad \qquad -2 \frac{c^2}{b^2} \left( \sin\psi\cos\theta\cos\phi + \cos\psi\sin\phi \right) \left( \cos\psi\cos\theta\cos\phi - \sin\psi\sin\phi \right) - \sin^2\theta\sin 2\psi  \Bigg] \notag \\
& \phantom{\equiv} + \frac{\sin^2\theta}{2} \Bigg\{ \left[ \sin\psi \cos\theta \left( \cos 2\phi \left( \frac{c^2}{b^2} - \frac{c^2}{a^2} \right) + \frac{c^2}{a^2} + \frac{c^2}{b^2} - 2 \right) + \cos\psi \sin 2\phi \left( \frac{c^2}{b^2} - \frac{c^2}{a^2} \right) \right] \notag \\
 & \qquad \qquad \qquad \times \left[ \sin\psi \sin 2\phi \left( \frac{c^2}{a^2}-\frac{c^2}{b^2} \right) - \cos\psi \cos\theta \left( \cos 2\phi \left( \frac{c^2}{a^2} - \frac{c^2}{b^2} \right) + 2 -\frac{c^2}{a^2} - \frac{c^2}{b^2} \right) \right] \Bigg\} \; , \notag \\
\notag \\
C & \equiv \left[ \sin^2\theta \left( \frac{c^2}{a^2} \sin^2\phi + \frac{c^2}{b^2} \cos^2\phi \right) + \cos^2\theta \right] \times \Bigg[ \frac{c^2}{a^2} \left( \cos\psi \cos\theta \sin\phi + \sin\psi \cos\phi \right)^2 \\ 
 & \qquad \qquad + \frac{c^2}{b^2} \left( \cos\psi \cos\theta \cos\phi - \sin\psi \sin\phi \right)^2 + \cos^2\psi \sin^2\theta  \Bigg]  \notag \\
 & \phantom{\equiv} - \frac{\sin^2\theta}{4} \left\{ \cos\psi \cos\theta \left[ \cos 2\phi \left( \frac{c^2}{a^2} - \frac{c^2}{b^2} \right) + 2 -\frac{c^2}{a^2} - \frac{c^2}{b^2} \right] + \sin\psi \sin 2\phi \left( \frac{c^2}{b^2} - \frac{c^2}{a^2} \right) \right\}^2 \; . \notag
\end{align}

\end{appendix}

\begin{appendix}
\section{A method for accelerating the computation of deflection angles}
\label{app:speed_up}

\citet{2003ApJ...599....7O} show that the surface mass density profiles of triaxial haloes are ellipsoidal. Let $q$ denote the corresponding minor-to-major axis ratio \citep[cf.][Eq. (35)]{2003ApJ...599....7O}. Then, the components of the deflection angle $\boldsymbol{\alpha} = (\alpha_x, \alpha_y)$ can be expressed as one-dimensional integrals of the convergence $\kappa(\zeta)$ \citep[see also][]{1990LNP...360...46S, 2001astro.ph..2341K}:

\begin{align}
\label{eq:alpha_x}
\alpha_x(x,y) = qxJ_1(x,y) \; , \\
\alpha_y(x,y) = qyJ_0(x,y) \; ,
\label{eq:alpha_y}
\end{align}
where the integrals $J_n(x,y)$ are defined as

\begin{align}
J_n(x,y) &= \int_0^1\frac{\kappa\left[\zeta(v)\right]}{\left[1-\left(1-q^2\right)v\right]^{n+1/2}} \; \mathrm{d}v \; , \\ 
\zeta^2(v) &= v\left[y^2 + \frac{x^2}{1-\left(1-q^2\right)v}\right] \; .
\label{eq:zeta_of_v}
\end{align}
The numerical evaluation of these integrals is computationally quite expensive. Since $\lim_{v \to 0} \zeta(v) = 0$ and $\lim_{\zeta \to 0} \kappa(\zeta) = \infty$ (cf. Eqs. (40) and (41) from \citet{2003ApJ...599....7O}), the integrand is singular at the lower integration boundary. Accordingly the numerical integration routines converge only slowly. We found that this inconvenient behaviour can be avoided by two subsequent substitutions. First, we invert Eq. \eqref{eq:zeta_of_v} and solve for $v$:

\begin{equation}
v(\zeta) =
\begin{cases} \dfrac{\zeta^2\left(1-q^2\right)+x^2+y^2\pm\sqrt{\left[\zeta^2\left(1-q^2\right)+x^2+y^2\right]^2-4\zeta^2y^2\left(1-q^2\right)}}{2y^2\left(1-q^2\right)} & \quad (y \neq 0) \\
\dfrac{\zeta^2}{x^2+\left(1-q^2\right)\zeta^2} & \quad (y = 0)
\end{cases}
\end{equation}
For $y \neq 0$, $\lim_{v \to 0} \zeta(v) = 0$ implies that the negative branch must be the correct solution for $\zeta \ll 1$. Since $\left\{\left[\zeta^2(1-q^2)+x^2+y^2\right]^2-4\zeta^2y^2(1-q^2)\right\}^{1/2}$ never vanishes for $y \neq 0$ and the solution must be continuous, the negative branch is the unique solution for the complete interval. Given this relation, we substitute $v$ by $\zeta(v)$ and obtain

\begin{align}
J_n(x,y) &= \int_0^1\frac{\kappa\left[\zeta(v)\right]}{\left[1-\left(1-q^2\right)v\right]^{n+1/2}} \; \mathrm{d}v \\
         &= \int_0^{\zeta(1)}\frac{\kappa(\zeta)}{\left[1-\left(1-q^2\right)v\left(\zeta\right)\right]^{n+1/2}} \; \frac{\mathrm{d}v}{\mathrm{d}\zeta} \; \mathrm{d}\zeta \; ,
\intertext{where}
\frac{\mathrm{d}v}{\mathrm{d}\zeta} &=
\begin{cases}
\dfrac{\zeta}{y^2}-\dfrac{\zeta\left[\left(1-q^2\right)\zeta^2+x^2+y^2\right]-2y^2}{y^2\left\{\left[\left(1-q^2\right)\zeta^2+x^2+y^2\right]^2-4\zeta^2y^2\left(1-q^2\right)\right\}^{1/2}} & \quad (y \neq 0) \\
 & \\
\dfrac{2\zeta x^2}{{\left[x^2+\left(1-q^2\right)\zeta^2\right]}^2} & \quad (y = 0)
\end{cases}
\; .
\intertext{The obvious singularity at the lower integration boundary vanishes after this transformation, but the integrand is not smooth at the origin yet. We overcome this problem by making the additional substitution $\zeta = t^2$, so that} 
J_n(x,y) &= \int_0^{\sqrt{\zeta(1)}}\frac{\kappa(t)}{\left[1-\left(1-q^2\right)v(t)\right]^{n+1/2}} \; 2t \; \frac{\mathrm{d}v}{\mathrm{d}\zeta} \; \mathrm{d}t \; .
\end{align}
These simple integral transformations speed up our computations by a factor of $\sim 10$.

\end{appendix}

\begin{appendix}
\section{Sampling cosmological populations of dark matter haloes}
\label{app:drawing_cosmological_populations}

In several sections of this paper, we needed to sample realistic catalogues of dark matter haloes, where 'realistic' means that they should have similar mass spectra as halo populations in numerical simulations and the observed universe. For that purpose, we adopted the following common Monte Carlo approach \citep[see][for instance]{2009MNRAS.392..930O}, which is based on the Press-Schechter mass function \citep{1974ApJ...187..425P}. Let us assume we need to draw a random population of dark matter haloes within the mass range $[M_{\mathrm{min}},M_{\mathrm{max}}]$ and the redshift interval $[z_{\mathrm{min}}, z_{\mathrm{max}}]$. Then, we first subdivide both ranges into smaller, equidistant bins, adopting typical bin sizes of $\Delta z = 0.02$ and $\Delta(\log(M)) = 0.02$. Next, we calculate the mean expected number $\bar{N}$ of haloes in each bin using the Press-Schechter mass function $\mathrm{d}n(M,z)/\mathrm{d}M$,
\begin{equation}
\bar{N} = \frac{\mathrm{d}^2N}{\mathrm{d}z \: \mathrm{d}M} \; \Delta z \, \Delta M = \frac{\mathrm{d}n(M,z)}{\mathrm{d}M} \, \Delta M \times \frac{\mathrm{d}V}{\mathrm{d}z} \, \Delta z \; ,
\end{equation}
where $\mathrm{d}V/\mathrm{d}z$ denotes the differential comoving volume. After that, we generate a random integer number $N$ from the Poisson distribution with mean $\bar{N}$. Finally, we sample $N$ haloes within the current bin and assign them uniformly distributed masses and redshifts out of the corresponding ranges. 

This Monte Carlo method is simple and efficient. It reliably generates random halo catalogues whose deviations from the Press-Schechter mass function remain within the limits of Possion noise.

\end{appendix}

\end{document}